\documentclass[conference]{IEEEtran}
\usepackage[a4paper,left=3cm,right=2cm,top=2.5cm,bottom=2.5cm]{geometry}
\usepackage{cite}
\usepackage[pdftex]{graphicx}
\usepackage[cmex10]{amsmath}
\usepackage{amssymb}

\usepackage{times}
\usepackage{array}
\usepackage[tight,footnotesize]{subfigure}
\usepackage[font=footnotesize]{subfig}
\usepackage{graphicx}
\usepackage{mathtools}
\usepackage{amsmath}

\usepackage{alltt}

\def\boxx{{\vcenter{\vbox{\hrule height.3pt
          \hbox{\vrule width.3pt height6pt
          \kern6pt\vrule width.3pt}\hrule height.3pt}}\;}}

\def\impos{{\;\vcenter{\hbox{\rule[0.15mm]{1.8mm}{1.7mm}}} \;}}

\def\lrarrow{\leftrightarrow \kern-8pt \rightarrow}

\def\2{\frac{1}{2}}

\def\beq{\begin{eqnarray}}
\def\eeq{\end{eqnarray}}
\def\2{\frac{1}{2}}
\def\AND{\,\wedge\,}

\newtheorem{assumption}{Assumption}

\newtheorem{example}{Example}
\newtheorem{lemma}{Lemma}

\def\tick{\text{\sc tick} }
\def\tock{\text{\sc tock} }
\def\Lteck{\text{\sc L-teck} }
\def\Rteck{\text{\sc R-teck} }
\def\Ltack{\text{\sc L-tack} }

\def\Rtack{\text{\sc R-tack} }
\def\Ltick{\text{\sc L-tick} }
\def\Rtick{\text{\sc R-tick} }
\def\teck{\text{\sc teck} }
\def\tack{\text{\sc tack} }
\def\nack{\text{\sc nack} }

\def\lrarrow{\leftrightarrow \kern-8pt \rightarrow}

\def\frightarrow{\rightarrow \kern-11pt /~~}
\def\reducesto{\simeq \kern -3pt >}

\usepackage{fancyhdr}					
\begin{document}
\newcommand{\strust}[1]{\stackrel{\tau:#1}{\longrightarrow}}
\newcommand{\trust}[1]{\stackrel{#1}{{\rm\bf ~Trusts~}}}
\newcommand{\promise}[1]{\xrightarrow{#1}}
\newcommand{\revpromise}[1]{\xleftarrow{#1} }
\newcommand{\assoc}[1]{{\xrightharpoondown{#1}} }
\newcommand{\rassoc}[1]{{\xleftharpoondown{#1}} }
\newcommand{\imposition}[1]{\xrightarrow{#1} \kern-10pt \impos}

\newcommand{\scopepromise}[2]{\xrightarrow[#2]{#1}}
\newcommand{\handshake}[1]{\xleftrightarrow{#1} \kern-8pt \xrightarrow{} }
\newcommand{\cpromise}[1]{\stackrel{#1}{\frightarrow}}
\newcommand{\policy}{\stackrel{P}{\equiv}}
\newcommand{\field}[1]{\mathbf{#1}}
\newcommand{\bundle}[1]{\stackrel{#1}{\Longrightarrow}}


\title{Spacetime-Entangled Networks (I)\\\Large Relativity and Observability of Stepwise Consensus}

\author{Paul Borrill, Mark Burgess, Alan Karp, and Atsushi Kasuya\\~\\Earth Computing Inc, CA, USA}
\maketitle
\IEEEpeerreviewmaketitle


\renewcommand{\arraystretch}{1.4}

\begin{abstract}
  Consensus protocols can be an effective tool for synchronizing small
  amounts of data over small regions.  We describe the concept and
  implementation of entangled links\cite{paulpatent}, applied to data
  transmission, using the framework of Promise Theory as a tool to
  help bring certainty to distributed consensus.  

  Entanglement describes co-dependent evolution of state.  Networks
  formed by entanglement of agents keep certain promises: they deliver
  sequential messages, end-to-end, in order, and with atomic
  confirmation of delivery to both ends of the link.  These properties
  can be used recursively to assure a hierarchy of conditional
  promises at any scale.  This is a useful property where a consensus
  of state or `common knowledge' is required. We intentionally
  straddle theory and implementation in this discussion.
\end{abstract}



\section{Introduction} 

We describe the agents, promises, and assessments involved in making a
multi-layer message-passing channel, using the property of
entanglement. Entangled network links are a concept for transmitting
data transactionally. They keep basic promises about message integrity.
Their design principles can be applied recursively to build a
transaction based hierarchy of communication, from the packet to the
application level.

Entanglement networks, proposed by Borrill\cite{paulpatent}, promise
to deliver sequential messages, end-to-end, in a path-independent
invariant order, and with confirmation of delivery to both ends of the
link. These are properties more usually associated with database
transactions.  While this may seem trivial over a point to point
link, it is a useful low level property for building cases where a
consensus of state, or common knowledge, is required.  Some of the
properties mentioned are available in any {\em reliable} message
protocol (e.g.  TCP), but in a form that may not be
optimal\cite{BorrilBCD14}.  In this work, we establish the hierarchy
of constraints for keeping low level promises about reliable message
propagation.

\subsection{Entanglement}

Entanglement is an information theoretic property\cite{giacomo1},
which means that several agents act inseparably, i.e. their outcomes
are co-determined.  In pair-entanglement, whatever is promised by one
end of the link, depends on what is (or was) promised by the other end
of the link in an absolute and causal manner, and vice versa.  This
constraint of mutual circularity (a form of semantic `deadlock' in
computer science) has eigenstates that admit strongly correlated or
anti-correlated behaviours, and may be used to lock the endpoints into
acting as a unit. Correlated behaviour allows communicating parties to
keep in step.  {\em Anti}-correlated behaviour
allows the end points to distinguish their separate identities without
ambiguity, so that input and output are expelled to the extrema of the
link, away from the middle, where non-intentional noise sources could
interfere. This suggests that the intentional management of boundary
conditions alone could be used to modulate a state, and propagate a
message, making the channel quasi-synchronous and quasi-deterministic.
We shall explain the essence of these claims below.

If entanglement can be maintained, over successive
promise-keeping exchanges, then, even as data are passed across the
link, a kind of temporal integrity (stepwise ordered delivery without dropped
packets) could be maintained. The scaling of the concept is the
key to its applicability: our aim is to make it an irreducible
promise, kept by an effectively irreducible link, acting as a single
scaled agent\cite{spacetime2}, meaning that one end of a link cannot
act without consequences for the other end.  It is this irreducibility
that one hopes to exploit in designing a mechanism for reliable
communication at the most primitive level.

The paper has the following structure:
\begin{itemize}
\item We define the concepts and notations.
\item We define senders, receivers, messages and packets.
\item Messages and packet encodings are defined in some detail.
\item Implementation of a link entanglement is shown.
\item We assess some of the promises of entanglement 
relative to alternative technologies.
\end{itemize}

\subsection{Motivation for a new delivery mechanism}

Conventional link layer protocols are `unreliable', meaning that they
make no assurances of delivery\cite{tannenbaum1,BorrilBCD14}.  Higher
level constructs, such as TCP reconstruct reliability with
acknowledgement circuits, but the application streaming abstraction
limits one's ability to reason about failures, and pushes recovery to
the application layer: if a packet does not arrive within a certain
time, one is not sure why or what to expect. Many failure modes in
communication are related to infrastructure issues, which could
reasonably be expected to self-repair. However, useful feedback is
lacking at the infrastructure level, because the reliability circuitry
is at too high a level. Our proposal here is to push the feedback
circuitry down the stack in a fundamental way.

In telephony, lower layer protocols like the ITU's Public Network
Signalling SS7 protocol, MTP layers 2 and 3 offer lower level
reliability and recovery quite similar to the scheme described
here\cite{ss7}.  Of the many others, we mention ATM\cite{atm}, Virtual
Synchrony\cite{virtualsynchrony}, Distributed State Machine
Replication\cite{rdma}, etc.  However, these protocols work atop
opportunistic signalling mechanisms, and lack the ability to say
precisely when a packet was delivered, from one end to another, at
each leg of a journey. This resilience properties have to be built top
down, instead of bottom up, where they can have a more effective
impact.  If one cares to know when data were received, e.g. during
financial transactions, or when parties knew certain facts within a
chain of evidence, then both sides of a transfer can benefit from
finer grained control over transactions, working together with a
smarter automated infrastructure.

The key underlying limitation on predictability is the indeterminism of
network communications, compared say to the relative determinism of
communication of a motherboard PCI bus. We pose the question: could one
construct the same level of deterministic trust on a cloud scale, as
for a single motherboard computer bus. Scaling a set of
deterministically reliable promises is a challenge that deserves
serious consideration, especially in the context of `cloud' or utility
computing.

The infamous Fischer, Lynch and Patterson (FLP) result\cite{flp},
which details how consensus is impossible in an asynchronous system if
agents only {\em might} be unreliable, is one of the most discussed
issues in system reliability. Consensus
protocols\cite{paxos,raft,bitcoin} make a variety of promises
concerning transactional ordering and versioning races.  Entangled
links, as described here, might be helpful in calibrating what data
units consensus might profitably be applied to. One can engineer
`observability controls' into the approach, limiting or locking what
observers can see at each moment. Endpoints recover synchronicity in
communications by being exempted from low level details, and
preventing contentious races from flapping data unnecessarily.
Conceptually, by altering the way states change (or whimsically by
redefining the way time is counted) at the communicating endpoints,
the parties in a system could be prevented from seeing inconsistent
states before clear promises can be made about their interpretation.
The `management of moments' might be applied at a number of levels.
In this note, we focus on explaining the mechanism for general
application.

\subsection{Promises}

What promises do agents need to make at each scale to claim such
properties?  Ordered delivery is not so much the issue: within an
identifiable message, packets can and usually have to be numbered, so
this is not a serious problem.  However, for messages with
disconnected origins, such as new and competing transactions to a bank
account, asserted as impositions\cite{TIP} rather than being promised
by mutual arrangement, conflicts can arise, and there is uncertainty
about when or whether packets were delivered.  Packets that arise from
independent sources and locations cannot be ordered
meaningfully\footnote{For some purposes one can approximately
  calibrate a set of timestamps by approximately synchronizing clocks
  through a single time source, assuming that they run at
  approximately the same rate, under approximately the same
  conditions, but this is liable to run into problems where the
  approximations fail.} with respect to one another at the sources,
since the sources have no calibrated causal alignment, rather it is
important to know in what order they were accepted by an aggregator.

Causal co-dependence is the common theme in this work.  Whatever the
agents in a distributed system promise, they must promise it together,
as an indivisible causal unit in order to keep in step with one
another. That umbrella promise may be used to engineer a perception of
determinism, and to place precise limits on what is meant by
simultaneity inside the system.  Ultimately, the goal is to be able to reason
deterministically about states (with all the advantages that entails),
rather than handing off responsibility to application level logic, which
is already impaired by the uncertainty of the layers underneath.

This, then, is a reasonable foundation for certainty.  Entanglement is
not a transitive property, so wide area communications cannot assume
the same promises as agents over a small scale, but one expects that
approximations can be worked, by forwarding through trusted
intermediaries, or the recursive application of the entanglement
method. These methods could make particular sense in environments with
physical assurances, like datacentres, and enclosed circuitry.  At
first glance, entangled links might not seem to contribute anything
new in the field on networking, but the benefits can be expected to
lie mainly in pursuit of the goal of `knowability' in a
space characteristically beset by uncertainties.

\section{Notation and configuration}

The idea of entangled links is quite simple, but turning the idea
into a message passing channel involves a surprising number of details.
In this section, we establish a language and notation for describing the parts of
the system. The basic arrangement is shown in figure \ref{link}.

\subsection{Cell agents}

The independent locations of interest in a network (i.e. the promise
theoretic intentional agents) are called {\em cells}. They are the
originators (sources) and recipients (sinks) of messages in a system;
e.g.  they may be `servers' in a datacentre, but they could also be
virtual units of agency, such as an application container or a sensor.
They must have a certain amount of memory for recording state, and
buffering messages decomposed into packets (a message is defined to be
a collection of packets).  We denote them by $C_i$ (where
$i=1,2,3,\ldots$ runs over all cell agents). Cells can encapsulate any
number of layers of entanglement in order to pass aggregations of
atomic messages across aggregations of intermediate agents,
hierarchically; however, we shall only illustrate the principles using
two layers in this paper.

We begin, therefore, with the interaction of a single pair of cell
agents, connected by a dedicated network channel. We label these $C_L$
and $C_R$ in their roles as `left' and `right' ends of a connection.
These will variously play the roles of sender and receiver.  The cell
agents' role in this discussion in to maintain state on the level of a
message, and to preserve a memory of their own identity during the
entanglement at the level of messages.  They must also promise
sufficient memory for message queues (buffers) of some length, for
incoming and outgoing messages; no message can have meaning without
memory to hold it, as an entity, in its entirety.
These assumptions prove to be of central importance.

\subsection{Network interface agents}

Two nodes, which we may arbitrarily refer to as left $C_L$ and right
$C_R$ communicate via a network channel whose endpoints are the
network interfaces $N_{L}$ and $N_{R}$ respectively.  Network
interfaces are intermediary proxies for cells they attach to. They are
promise-theoretic agents, since they make specific localizable
promises.  Interface agents are much simpler agents than cells, with
only registers for sending and receiving single packets. Their
capabilities determine the promises they can keep (and vice versa),
they are one-to-one\footnote{We shall discuss a couple of alternative
  ways in which the interfaces can keep promises, suitable for
  deterministic transmission, with more efficient promise keeping at
  the expense of additional `pipeline' state-memory in the interface
  agents, to avoid wasting cycles of communication while confirming
  packets already sent.}.

Each cell agent `contains' (or is associated with\cite{spacetime2})
one or more network interfaces, going to different neighbouring
destinations\footnote{What this containment means promise
  theoretically remains to be explained. See \cite{spacetime2} for a
  detailed discussion.}.  How we de-mark the boundary between cell and
network interface is not a uniquely defined matter, but it is
important to the matter of how we restrict the observablity of
intermediate and uncertain states, in order to bring certainty.
We lay out some conventions below.

We shall assume that a network interface promises only packet
delivery, not extended message delivery: message delivery is for other
agencies with the cells to ensure, building on the promises of
reliable atomic packet delivery (in the manner of an OSI layer model).

Network interface agents are the promisers in the mechanics of data
transmission and entanglement.  Each agent $C_{i}$ thus has a number
of network interfaces $N_j$, which are formally independent of
$C_i$.  Each network interface agent has registers fulfilling two
roles: for sending and for receiving data intended for its adjacent
neighbour counterpart.  Network interfaces act as dispatchers, for
the serial queue used by each cell for message passing.

The $N_i$ network interfaces act as proxies for the cells $C_i$,
making dedicated `dumb' promises, subordinate to the message queue
imposed by the cell. The promises between $N_i$ and $C_i$ are thus crucial for
building higher levels of entanglement on top of lower layers (figure \ref{entangled}).
We introduce two new agents: $Q_L$ and $Q_R$ for the message queues that feed into
and out of the link.

\subsection{Network registers}

Each network interface possesses effectively two logical
ports (i.e. two registers): one for `data out' or sending called
$N_i^{(+)}$, and one for `data in' or receiving called $N_i^{(-)}$,
where $n = 1,2,3, \ldots$ labels a particular network interface,
attached to any cell node.  Network interfaces are formally separate
agents from the cells (as required in promise theory, since they
maintain independent promises); therefore, we must eventually describe
the cooperative promises between the cells and their interfaces too. However, we
focus mainly on the network interface agents for now.

\begin{figure}[ht]
\begin{center}
\includegraphics[width=7.5cm]{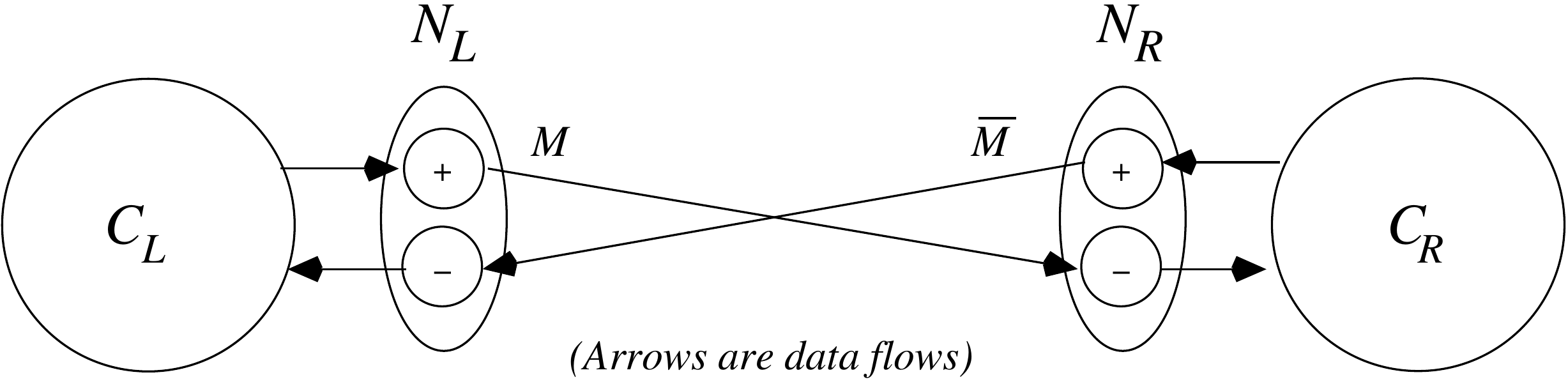}
\caption{\small The agent structure of a link, and the direction of data flows (not promises) between them.
$C_L$ and $C_R$ may be sender or receiver. Network interface agents
for these are denoted $N_L$ and $N_R$, with
internal register agents $N^{(+)}$ and $N^{(-)}$ for sending and receiving respectively.
A message sent from the role of sender to receiver is denoted $M$, and the complementary reverse 
acknowledgment message is denoted $\overline M$.
Note that, by promise theory principles, we deal with the communication between
each pair of agents as a separate set of promises, since the behaviour of each agent is independent.
\label{link}}
\end{center}
\end{figure}

\begin{figure}[ht]
\begin{center}
\includegraphics[width=6.5cm]{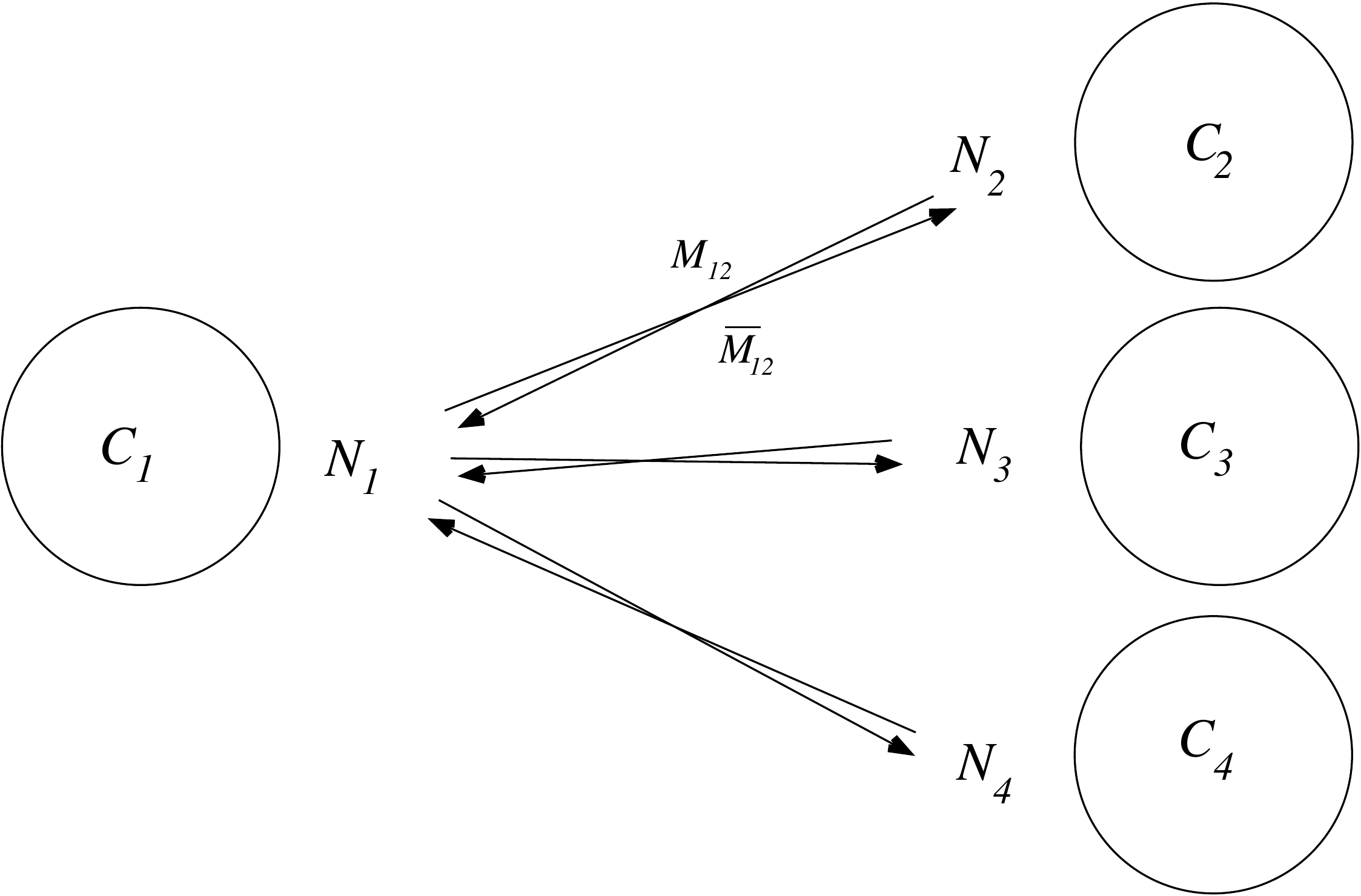}
\caption{\small More cells, with interfaces, and their labels. With
  multiple interfaces per cell, the labels left/right are no longer a
  good designation. We use them in this paper for consistency.\label{links}}
\end{center}
\end{figure}

Let any network interface agent $N_i$, $i = 1,2,3,\ldots$ be denoted by
a doublet of two registers: $N^{(+)}_i$ for sending, and $N^{(-)}_i$ for receiving: 
\beq
N_i = \left(
\begin{array}{c}
 { N}^{(+)}_i\\ 
 { N}^{(-)}_i
\end{array}
\right)
= \left(
\begin{array}{c}
 \text{send register}\\ 
 \text{receive register}
\end{array}
\right) 
\eeq 
as in figure \ref{link}.  
Figure \ref{links} clarifies the notation or more than two interfaces,
distributed across a  number of cell agents.

\subsection{Geometry of the configuration (physical symmetry breaking)}

\begin{figure}[ht]
\begin{center}
\includegraphics[width=6cm]{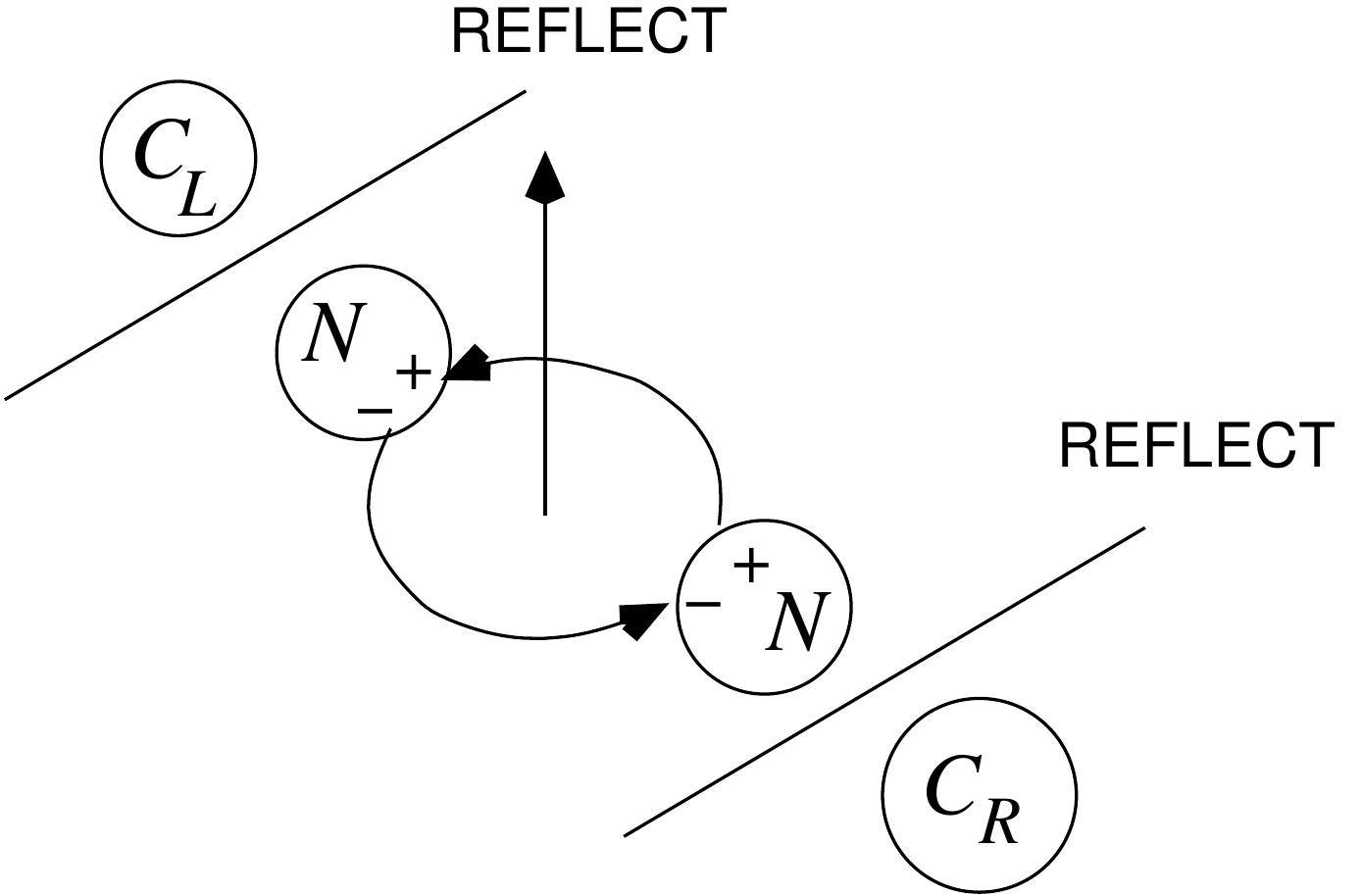}
\caption{\small If the two directions of message travel between left
  and right pass along separate channels, the orientation of the
  loop they form also has an orientation.\label{oriented}}
\end{center}
\end{figure}

The geometry of a link is shown in figure \ref{oriented}, with left
and right ends of an axial link. The operation of reflection may be
interpreted digitally as a NOT operation\footnote{In the original
  document, the exchange of vector items was used; such an exchange is
  a $n+1$ dimensional linear representation of a NOT operation in $n$
  dimensions.}.  If the two directions of message travel between left
and right pass along separate channels, i.e. full duplex (see figure
\ref{oriented}), the orientation of the loop they form also has a
topological orientation\footnote{The direction of the current define
  something analogous to a magnetic field direction for the loop.}.
As messages pass back and forth they have two eigenmodes: the digital
equivalent of standing waves (which we shall use for equilibrium
states, and travelling waves, which we shall use to transmit data
packets.

\subsection{Bootstrap of $L,R$ orientations (dynamical symmetry breaking)}

Network interfaces, connected for the first time by a network channel,
can self-organize to select an arbitrary naming convention or
`orientation' for the endpoints. A protocol flips a virtual coin to
break the symmetry between left and right, and decide which agent can
promise to call itself `left' and which promises to call itself
`right'.  Agents converge on this state of broken
symmetry\footnote{This is like a trivially local ferromagnetic state
  along each interface-to-interface change, but with no long range
  correlation between different interface pairs.}.  Thus, they do not
need to be assigned endpoint addresses by an exterior authority:
geometry alone assigns them designations of `self' and
`non-self'\footnote{This notion of `address-less endpoints' (meaning
  point to point links in which the addresses are redundant by virtue
  of geometry) is used in another promise oriented system: unnumbered
  interfaces in BGP. In a point to point link, the link geometry can
  prevent misunderstandings about intended source and destination:
  each endpoint recognizing self and non-self.}; one only needs to
distinguish one end from the other in a mutually agreed way.  We use
the convention $L$ and $R$ for `left' and `right' (non-left or
$\overline{\rm left}$, as opposed to heads/tails, up/down, black/white
etc) for these designations.

By convention, the left agent will send signals \tick and the right agent
replies with \tock, in the language of \cite{classent}. We shall also use the terms
$\Ltick$ and $\Rtick$ for consistency with the unified narrative about symmetry
breaking and orientation of the link. 

These designations promising $L$ and $R$ are made by the network
interface agents, at the level of packets, and apply only to the
link\footnote{Although, for our two agent example, we use the cell
  designations $C_L$ and $C_R$, the cells have no natural left or
  right designation, as they have multiple link dimensions.}. It is
nonetheless necessary to maintain corresponding distinctions for
cells via the intended addressee of messages, e.g. to ensure that a
message doesn't accidentally get reflected back to its sender,
especially when travelling over long distances. This designation
cannot be inherited from the interface card, because a cell has
multiple network interfaces. Cells therefore need to refer to one
another in the scope of a larger model, which addresses the concerns of
messages.

\subsection{Intent versus timeless average behaviour}

After the initial negotiation of left-right symmetry breaking, to
establish the configuration in figure \ref{oriented}, and determine a
coordinate basis: $L$ who signals with \tick (or $\Ltick$) and $R$ who signals \tock
(or \Rtick), etc, 
two modes of operation can be defined across the link:
\begin{enumerate}
\item {\bf Intent free}: a idling or standby phase.
  This phase may be called `timeless', because no direction for
  advancement is selected that marks out a unidirectional timelike
  progression; the system goes as much forwards as it goes backwards,
  maintaining a steady state equilibrium. Exterior observers see no
  observable change in the promised state of the link or its
  endpoints, i.e.  no time passes on the link's exterior clock, and on
  the interior every forward step is met with a step backwards.

\item {\bf Intentional}: a transitory asymmetric message transmission
  phase. In this phase, the $L-R$ symmetry is broken, and observers
  exterior to the link see time advance one tick for each packet sent,
  for the duration of a message. Any (even) number of non-observable
  interior exchanges may need to be enacted in order to make this
  transfer of information happen.

\end{enumerate}
Entanglement persists, uninterrupted, through both modes of operation.
These two modes are referred to as ENTL and ENTT respectively in the
nomenclature of \cite{classent}.  Throughout both modes, the pendulum
nature of exchange continues verifiably, alternately from left to
right, and back.

\subsection{Two kinds of symmetry}

We refer to two separate symmetries in our description of the link system:
\begin{enumerate}
\item Spatial configuration symmetry (layout, characterized by $L,R$).
\item Dynamical balance symmetry, relative to (1), characterized by
  sender/receiver role exchanges $S,R$ or $+,-$.
\end{enumerate}
As a purely technical note, the left-right configuration handedness of
each link is not significant over long times i.e. over many
interactions, when at steady state station-keeping. It is averaged out
by dynamical interactions. In this sense, over arbitrary (even)
numbers of interactions, the link may be called timeless (like a
standing wave), since it is a superposition of two simple Markov
processes: it has no interior state that can act as memory, counting
like a clock of order greater than one. It is basically a pendulum
that cannot be observed microscopically. Its average position is zero,
neither left nor right.  Individual transmissions, piggy-backed over
this pendulum have a handedness (chirality), but even these may
average to zero if one disregards their semantic content.

\begin{example}
A simple analogy would be to think of a grandfather clock. The
orientation of the clock does not affect its ability to tell the time:
its pendulum does not swing more in one direction than the other, and
it doesn't matter which side we call left or right. Nevertheless, when
gears are engaged, the continuous motion drives an asymmetric
clockface in a single direction, by breaking the symmetry.  The
clockface is like a message being transmitted in one direction.
The pendulum (as a Markov process) cannot count or remember any length of
time, beyond one tick, but it can drive the transmission of a larger
sequence (on the clock face) to keep a time message, modulo the memory size of
the clock face.  In our example, each network interface is a pendulum,
and the clock face is analogous to a cell's private buffer, for
transmitting a message.
\end{example}

\begin{itemize}
\item When agents have no intent to direct a message, average equilibrium
  symmetry presides with exchanges of pendulum packet labelled \tick
  and \tock (heartbeats), which can be called timeless.
\item When agents intend to propagate a message in a single direction,
  they use the signals \teck (offer) and \tack (acknowledgment),
  generating ticks that mark an advancement of time at the end points.
\end{itemize}

\begin{figure}[ht]
\begin{center}
\includegraphics[width=4cm]{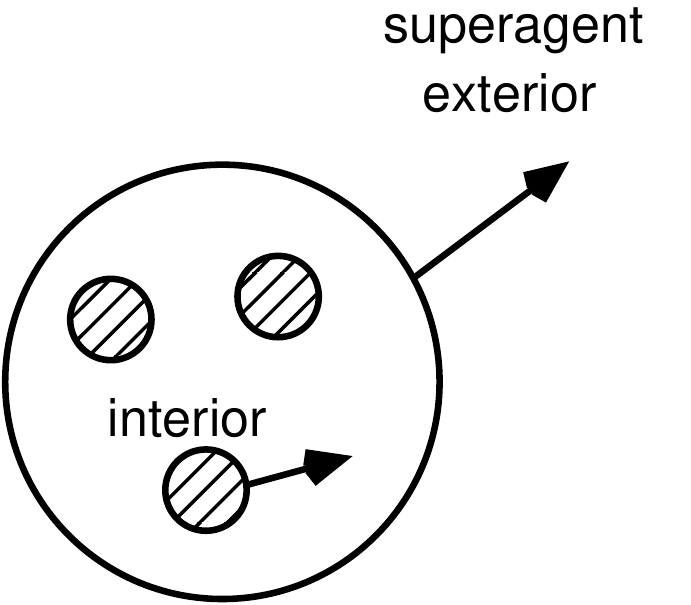}
\caption{\small An aggregation of agents working collectively
defines an effective entity or `superagent', which acts
as one, providing a level of encapsulation.
The agent is irreducible, if the agents within are
co-dependently bound by interior promises.\label{superagent}}
\end{center}
\end{figure}

When scaling these methods to aggregate `superagents' (see figure
\ref{superagent}), the same two phases are needed at each layer over
dependency that combines agent to agent entangled links into higher
level abstractions with the same irreducible semantics.  Thus we shall
refer to these two phases of exchange repeatedly: an equilibrium
(steady state) phase, and an intentional disequilibrium (transition)
phase\footnote{This two mode solution has a direct analogue in the
  solution of any set of constraints, e.g. in the solution of
  differential equations, with source term. There one has a
  `particular integral' or transitory response to specific boundary
  conditions, which dies out away from the source. The `complementary
  solution' or steady state equilibrium behaviour represents the
  average behaviour over long times (which is therefore relatively
  `timeless' compared to the timescale of the transitory changes). In
  our case, cell agents impose boundary conditions by the `intent to
  send'. When this response has played out, the idling steady state
  (timeless) behaviour persists.}.

\subsection{Encapsulation of exterior messages}

How we `quantize', or define the atomicity of outcomes, frames the way
in which we interpret the units of transfer in message delivery.
Entanglement (or irreducibility) of communication allows us to convert
non-deterministic asynchronous message channels into effectively
deterministic synchronous message channels, by restricting or `quantizing'
observability. This is a sleight of hand, based on the voluntary
cooperation of the agents involved, but it can be effective.

The property of entanglement has the consequence that, once a message
enters a link, it must either leave it as an indivisible unit, or have
no effect whatsoever, and thus each transferrable unit must be wholly
containable in the send and receive registers. No observer could see a
partial state.  Everything entering becomes a state of the collective
superagent.  In the language of distributed consensus, we can turn
these promises into commitments, kept deterministically for single
hops, and then build on the increased certainty to work towards larger
scale consensus\cite{paxos,twophasecommit,raft,promisebook}.
\begin{figure*}[ht]
\begin{center}
\includegraphics[width=16cm]{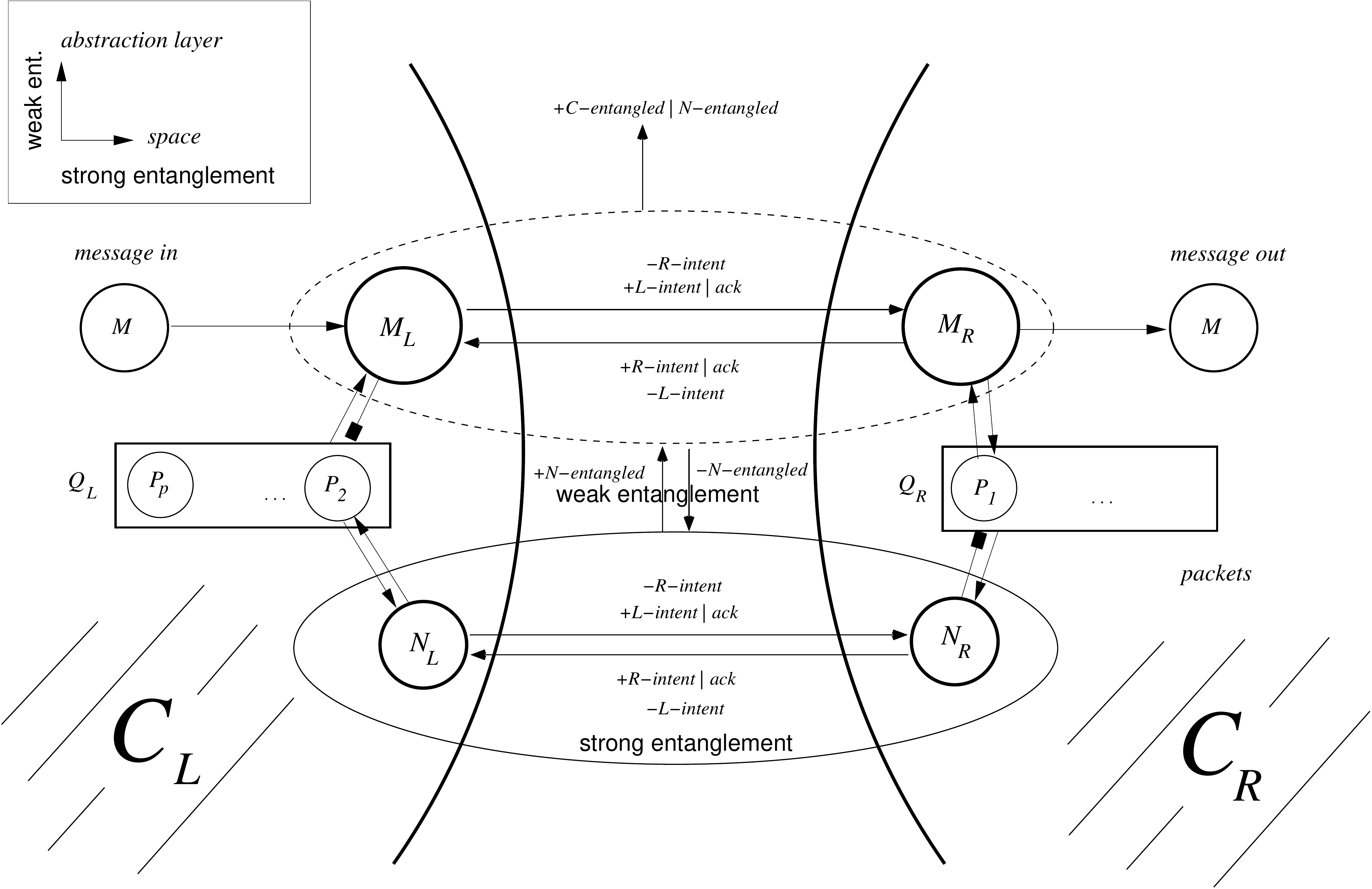}
\caption{\small Entanglement is a bottom-up property, where higher
  level promises depend on the entanglement of the lower layers (see section \ref{thepromises}). The
  entanglement makes pairs of agents effectively into a single
  irreducible superagent, in the sense of \cite{spacetime2}. By
  building on such dependencies, we can trade ad hoc homogeneity for large scale
  quasi-deterministic fabrics.\label{entangled}}
\end{center}
\end{figure*}

We packetize messages, as is normal to keep transactions predictable,
and thus at least two layers are needed for packetized message
delivery model (see figure \ref{entangled}): a network interface and
link layer operates on the level of packets, and an `intended message'
layer, for aggregating packets into larger messages, operates between
the cells.  The cells are the intentional agents, originating and
consuming messages as part of their larger plan. Network interface
agents have intent only by proxy. 

Separate entanglements may be established at each level of information
promises: it is the entanglement of the network interfaces that allows
us to make strong statements about transmission of packet chunks, and
the entanglement of the cell queues or application buffers, containing
the entire messages, which may synchronize complete ordered messages.

\section{Message structure}

We divide the discussion into two parts: how data are exchanged across
the link, and how messages are packetized and reassembled.  Once data
have been passed across the link, what the receiver does with the
message is none of the link's business, and thus remains undefined for
all scales larger than the region formed by the entangled agents.
This reflects the essential autonomy principle of promise theory.
Agents, on any scale, can distort or dump data, as enshrined in the
law of information integrity (see 7.2.2 of \cite{promisebook}); 
the purpose of entanglement is to enable the detection of these events.

We shall focus on cells and network interfaces one at a time, from the
bottom up, understanding that a close collaboration between the layers
is needed for the higher layers to function. For simplicity, we describing
the special case of one network interface per cell (a single link). In a 3 dimensional
fabric one would expect cells to have four to six neighbours, all with point to point
entanglements.

\subsection{Interior message-packet promises (irreducibility)}

Data packets transmitted from cell to cell are also formally another
kind of agent that make promises about encapsulation of data payload; they
are emitted and absorbed by cells that define what we mean by `fixed'
locations in a network\cite{spacetime1,spacetime2}; the structures
promised by messages, a message ID. A message that is considered
atomic at the level of cells, may have internal structure at the level
of the network interfaces: packets must promise which message they
belong to, with the message ID, and also have a message sequence
number corresponding to a fragment of the total payload.

A complete message is thus a doublet, i.e. makes two promises: \beq M^{(m)}
= (m,D^{(m)}), \eeq where $m$ is a unique message identifier, $D_m$ is
a data payload, e.g. $m = 1, 2, \ldots$.  We do not assume any
promises are made about interior structure of the data here.  

The packetization implies that messages are superagents,
composed of a sequence of packet agents.  Each packet promises to be a
member of the message ($m$), and hence carries both a message
identifier and a sequence number for the total ordering of packets
within $M^{(m)}$.

Let a message $M^{(m)}$, originating at cell agent $C_i$, be defined as an ordered set
of packets $P$:
\beq
M^{(m)} = \left\{ P^{(m)}_1, P^{(m)}_2, \ldots, P^{(m)}_p \right\},\label{mesg}
\eeq
for some sequence length $p$. We shall say that a message $M^{(m)}$ has been transferred from
a cell agent $C_i$ to $C_j$ when each packet $P^{(m)}_a \in M^{(m)}$
has been emitted from $C_i$ and absorbed by $C_j$, in a congruent order.

Each packet is a tuple consisting of a header vector $\bf H$, which we shall represent below as a
3-vector (see section \ref{implemH}), and a data payload $D$, which is a scalar, and may be empty (when
no message needs to be sent):
\beq
P^{(m)}_p = ({\bf H}^{(m)}_p, D^{(m)}_p) 
\eeq 
where ${\bf H}_p$ is a header vector, and $D_p$ is the payload data
fragment.

The totally ordered aggregation of all packets belonging to a single
message $D^{(m)}_p$ is thus precisely equal to $D^{(m)}$, which we
might write whimsically (for the association of ideas
and mathematical interest only) as a kind of path ordered integral, oriented
along the intended narrative of the sequence:\footnote{This also
  helps to alert to the fact that such a path aggregation is non-local
  in the sense that the order originates from the intent of a
  non-local agent, and is reproduced by the subordination of autonomy
  to that remote source.}: 
\beq 
D^{(m)} = P \;\int_{p} D^{(m)}_p.  
\label{integrate} 
\eeq

If no messages are sent, a link merely tick-tocks along with no
average direction.  When a cell agent intends a message to be sent to
a neighbour, it inserts the message in a packetized form into a serial
queue, at one end of the link, which acts as a source. The
bidirectional symmetry of the link is now broken by the presence of
such a message, on the level of cells. Meanwhile, the link continues
in a purely reactive state, transferring single packets from register
to register, with no average directional intent. In other words,
all directionality arises from the boundary intent promised by cells.

\subsection{Sender and receiver roles, and registers}

Packets are transmitted from the sender register of one $N_i$ to the
receiver register of the adjacent network interface $N_j$: 
\beq
N_i^{(+)} \imposition{+P} N_j^{(-)}.  
\eeq 
The paths in figure
\ref{link} therefore cross over, or must be interleaved by
multiplexing. We do not need to define which method is used here.  In
both directions there will be transmissions with the role of ``intent
to send'' and ``intent to receive''.  So message headers signal:
\begin{quote}
header $\rightarrow$ role $\otimes$ direction
\end{quote}
The message packet headers thus promise both {\em intent} and
direction, for the packet layer.  A similar header must be promised at
all layers that entangle.  

Packets (called \tick/\Ltick or \tock/\Rtick) that represent idling or steady-state
exchange, represent an absence of intent (the interface agent signals
only that it is alive and treading water).  Packets called \teck and
\tack express asymmetric data transfer (send and receive).  We use the
symbols from \cite{promisebook} $\emptyset,+,-$ for these cases:

\small
\begin{tabular}{ccl}
($\emptyset$) & \tick & Idle, pendulum mode \\
($+$) & \teck & Offer / Send (payload)\\
($-$) & \tack / \nack& Accept / Reject (payload)
\end{tabular}
\normalsize

These three signal types occur in both left or right
varieties\footnote{In earlier notes \cite{classent}, the terms AITS
  and AITR were used for \teck and \tack respectively.}.  In all
cases, the non-idling intentional exchanges graft themselves onto a
single invariant pendulum process, by superposition: \beq
\teck &=& \tick + \Delta_+\\
\tack &=& \tick + \Delta_- \label{superpos} \eeq a bit like Tarzan
swinging from cell to cell on pendular vines.  A \nack message, which
is a negative acknowledgment or rejection `not ready to accept
payload' or `not ack', has the same encoding as \teck, but with
mandatory empty payload, so there is no ambiguity in
coding\footnote{Note that \nack is so named because its representation
  is simply NOT-\tack. Its function is to essentially run subtime
  backwards to undo a partial transaction. It should not be confused
  with other implementations of NACK elsewhere. In UK English, one might say that
  such rejected packets are consigned to the NACKer's yard!}. One
could also simply revert to a \tick mode with no ambiguity, but an
explicit signal has pedagogical value.

Combined with a single bit, which encodes the intended destination (or
message orientation $L2R$ or $R2L$), these promises may be announced
with three bits of header data.

\section{Irreducibility of entangled links}

In the previous section, we described the arena for communication:
the configuration of key agents and their promises. In this section,
we discuss what is means to be entangled, and the intended function.

Once the roles of left and right have been established, the oriented
pair of agents can collectively make promises that the two agents
working alone cannot make\footnote{The `left' and `right' roles should
  never need to be reestablished once set up, but some catastrophic
  external event (like a cable breakage) might force this
  renegotiation.}, provided we can limit the observability of interior
promised outcomes and present the collective exterior promises
as quasi-atomic.

\subsection{Entanglement and steady state}

When agents collaborate, or act cooperatively, it is natural to define
a `superagent' to label them as a collective entity\cite{spacetime2}.
When such a superagent, composed like $S = A_1 \oplus A_2$, makes promises that
cannot be attributed to or kept by either of its components $A_1$ or
$A_2$ alone, then we say the sub-agents are entangled, and we say
that the superagent is irreducible\cite{spacetime2}. This happens when
promises are mutually conditional.

\begin{lemma}[Entangled with respect to intent $I$]
  Two agents $C_L$ and $C_R$ are said to be entangled or irreducible
  if the superagent $C_L \oplus C_R$ enveloping both of them makes a
  promise that neither of the two agents can make alone. This can only
  happen if each agent makes promises conditionally on promises made
  by the other.  $\Box$
\end{lemma}
This definition is compatible with the definition of entanglement in
information theory\cite{giacomo1}.  For any promise bodies $I_L,I_R$,
the necessary and sufficient solution to this condition is given by
\beq
C_L &\promise{+I_L|I_R}& C_R\label{a1}\\
C_R &\promise{-I_L}  & C_L\label{a2}\\
C_R &\promise{+I_R|I_L}& C_L\label{a3}\\
C_L &\promise{-I_R} & C_R.\label{a4} 
\eeq 
The proof is trivial: both sides promise $I_i$ ($i=L,R$) with a
dependence on the promise $I_{\overline i}$ from the other, else they
would promise independently which would contradict the definition.  If
the agents do not promise the explicit dependence on the other in
(\ref{a2}) and (\ref{a4}), then (\ref{a1}) and (\ref{a3}) are not
complete promises, by the conditional promise law 6.2 of
\cite{promisebook}, that no dependent promise can be given without
accepting the dependent promise of the other, thus $C_L$ must accept
$I_R$ and vice versa.

When $I_L$ or $I_R$ changes, these promises may be thought of as
cyclicly generating an evolving sequence of preconditions, which
unfolds as a chain of transaction events, until an equilibrium is 
possibly reached.

\subsection{Interior and exterior time and observability}

Entangement is a co-dependent causal evolution of state; i.e. it works
in both directions `at the same time', so we must be careful what we
entangle, how `the same time' is defined, and how directionality is
arranged.  It affects $n$-clusters of agents, where $n > 1$.  Promise
theoretically, we can observe that there are implicit timescales as a
result of irreducible co-dependence being composed from atomic
elements:

If we define a timescale by $\Delta t^{(S)}$ at scale $S$, measured
according to the clock of an exterior godlike observer (figure
\ref{synchasynch}), with access to all information, then each tick
corresponds to a single promise-keeping event. The cells cannot
observe these events, which happen in between the ticks of their
`proper time' clocks, so we might call this ability to observe the
most detailed equilibrating events {\em subtime}\footnote{Anyone who
  has used a version control system understands subtime as all those
  moments observers of the document repository cannot see, that lead to
  what was committed in each observable version.}.

A complete cycle of entangled co-dependent causation leads to a
natural coarse-graining of time that corresponds to the aggregation of
interspatial events (two agents $L,R$ keeping $+,-$ promises to close
the cycle).  \beq \left.
\begin{array}{c}
\left.
\begin{array}{c}
\left.C_L \promise{+I_L|I_R} C_R \right\rbrace  ~~\Delta t^{(1)}\\
\left.C_R \promise{~~-I_L~~}   C_L \right\rbrace~~\Delta t^{(1)}\\
\end{array}\right\rbrace ~~\Delta t^{(2)}\\
~\\
\left.
\begin{array}{c}
\left.C_R \promise{+I_R|I_L} C_L \right\rbrace~~\Delta t^{(1)}\\
\left.C_L \promise{~~-I_R~~}  C_R \right\rbrace~~\Delta t^{(1)}
\end{array}\right\rbrace ~~ \Delta t^{(2)}
\end{array}\right\rbrace ~~ \Delta t^{(4)}.
\eeq 
Entanglement thus implies quantization of both space and time,
because nothing independent can happen in an entangled network, but we
can only observe entanglement on a coarse-grained
timescale\footnote{It does not rule out other forms of quantization at
  a smaller or larger scale.}.  If we refer to $\Delta t^{(4)}$ as
{\em exterior time} or co-time, and $\Delta t^{(S)}$ for $S<4$ as {\em
  interior time}, then we can call $\Delta t^{(1)}$ specifically {\em
  subtime}. It is purely local, and not observable by any other agent.
The promise of entanglement (codependence) is only observable at a
timescale $S \ge 4$. These basic points will inform the discussion of
a protocol by which can use entanglement to built a
quasi-deterministic communication channel.

The two time rates tick with the passage of the following agents,
assuming the sender $S$ is $L$:

\begin{center}
\begin{tabular}{|c|c|c|}
\hline
\sc Align & \sc Interior & \sc Exterior\\
\sc Role & \sc Symmetric & \sc Anti-symmetry\\
\sc Orientation & \sc clock tick & \sc clock tick\\
\hline
$L/S\rightarrow R$ & $\tick/\teck$ & +$\Delta P$\\
$R\rightarrow S/L$  & $\tock/\tack$ & -$\Delta P$\\
\hline
\end{tabular}
\end{center}

\subsection{Irreducible superagent picture}

\begin{figure}[ht]
\begin{center}
\includegraphics[width=4cm]{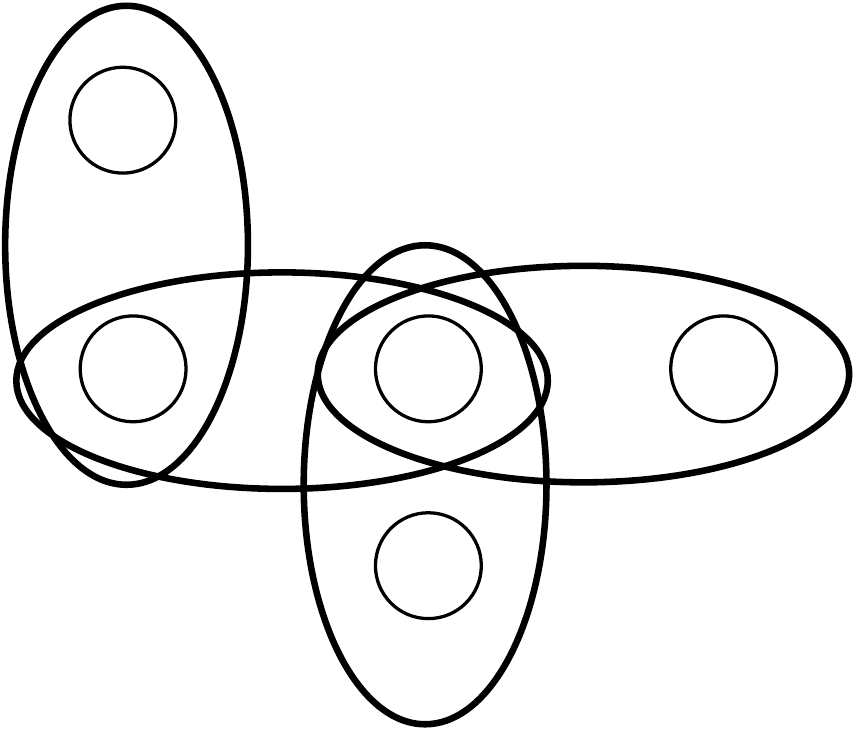}
\caption{\small Entanglement results in a new effective picture, with
  overlapping irreducible superagents. Entanglement (irreducibility)
  is not a transitive property, as the diagram shows: the overlapping
  of superagents does not imply a single large superagent keeping the
  same cooperative promises.\label{overlapsuper}}
\end{center}
\end{figure}

Co-dependent promises, made (and kept) by the endpoints, must be
maintained regardless of what other independent promises cells might
make to any other agent. This happens when both agents are driven by
what happens between them rather than coordinating their independent
activities (see figure \ref{overlapsuper}).  Our goal in this paper is
to explore the use of this property in order to keep strong promises
about message delivery.  Notice that these co-dependent promises are
invariant under $L \leftrightarrow R$, and are thus timeless and
without preferred orientation.

The keeping of this entangled state can be implemented in the
following signal promises, which may be considered an atomic cycle:
\beq
N_L &\promise{+\tick_{t+1}|\tock_t}& N_R\\
N_R &\promise{-\tick}& N_L\\
N_R &\promise{+\tock_{t}|\tick_{t+1}}& N_L\\
N_L &\promise{-\tock}& N_R 
\eeq 

Note also that this set is invariant under the symmetry $L
\leftrightarrow R$.  All these promises are invariant; they become
active and inactive based on the receipt of conditional signals.  The
repetition of this cycle of promises could be disturbed, in principle,
by the sending of a message to propagate data. However, we shall show
that the basic entanglement can be maintained even as data are
superposed on top of these promises, by defining superposed promises:
\beq
\teck &=& \tick + \Delta_+\\
\tack &=& \tick + \Delta_-
\eeq 
(see appendix equations
(\ref{sup1})-(\ref{sup2})).  To understand how this can help to maintain
certainty about the non-local state of the link, we need to explain
the non-local relativism of entanglement.

\begin{lemma}[Composition of entangled links]
The composition of irreducible or entangled links, as in figure \ref{overlapsuper}
cannot itself be irreducible or entangled. $\Box$
\end{lemma}
The proof of this follows from the linear combination omits off-diagonal
promises (see 11.1-13.1 in \cite{promisebook}).

\subsection{Single-valued time for paired agents}

\begin{figure}[ht]
\begin{center}
\includegraphics[width=6cm]{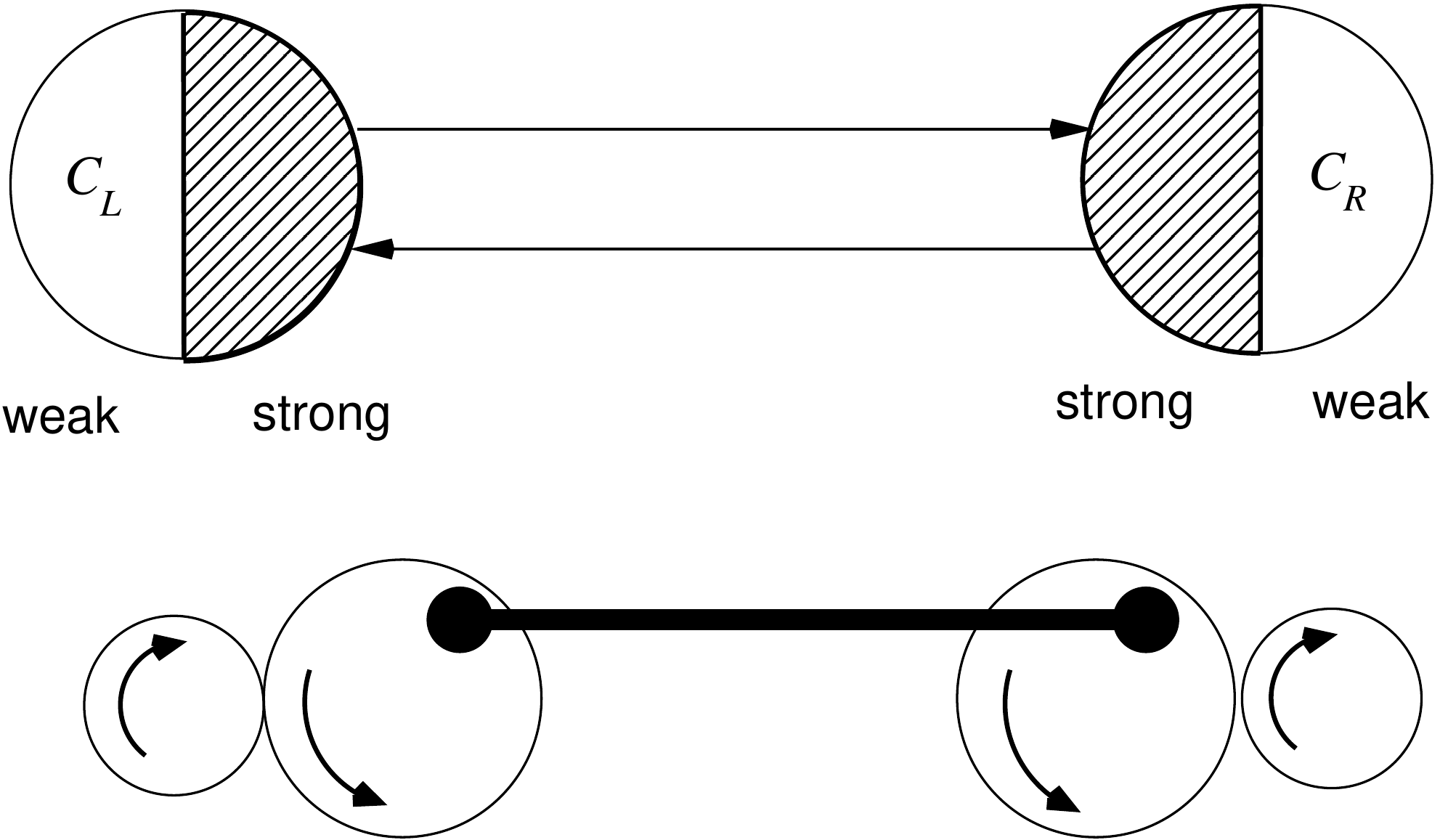}
\caption{\small Agents need to maintain a locally split brain model to
  retain control over the link, and avoid harmful deadlocks. If the
  link drives all aspects of the agents, they become too fragile, leading to
  possible failures of the link. This can be imagined as wheels joined
  rigidlt by a crank (in the entangled region), and decouplable gears
  that can be introduced to drive or be driven by the link.\label{sw}}
\end{center}
\end{figure}

In the geometry of the link, there are two distinct possibilities for
temporal evolution of the irreducible link superagent. We identify these as {\em
  local} and {\em non-local} in spacetime.  They correspond to how we define the clock by
which events move forward on the two ends of the link.
\begin{figure}[ht]
\begin{center}
\includegraphics[width=7cm]{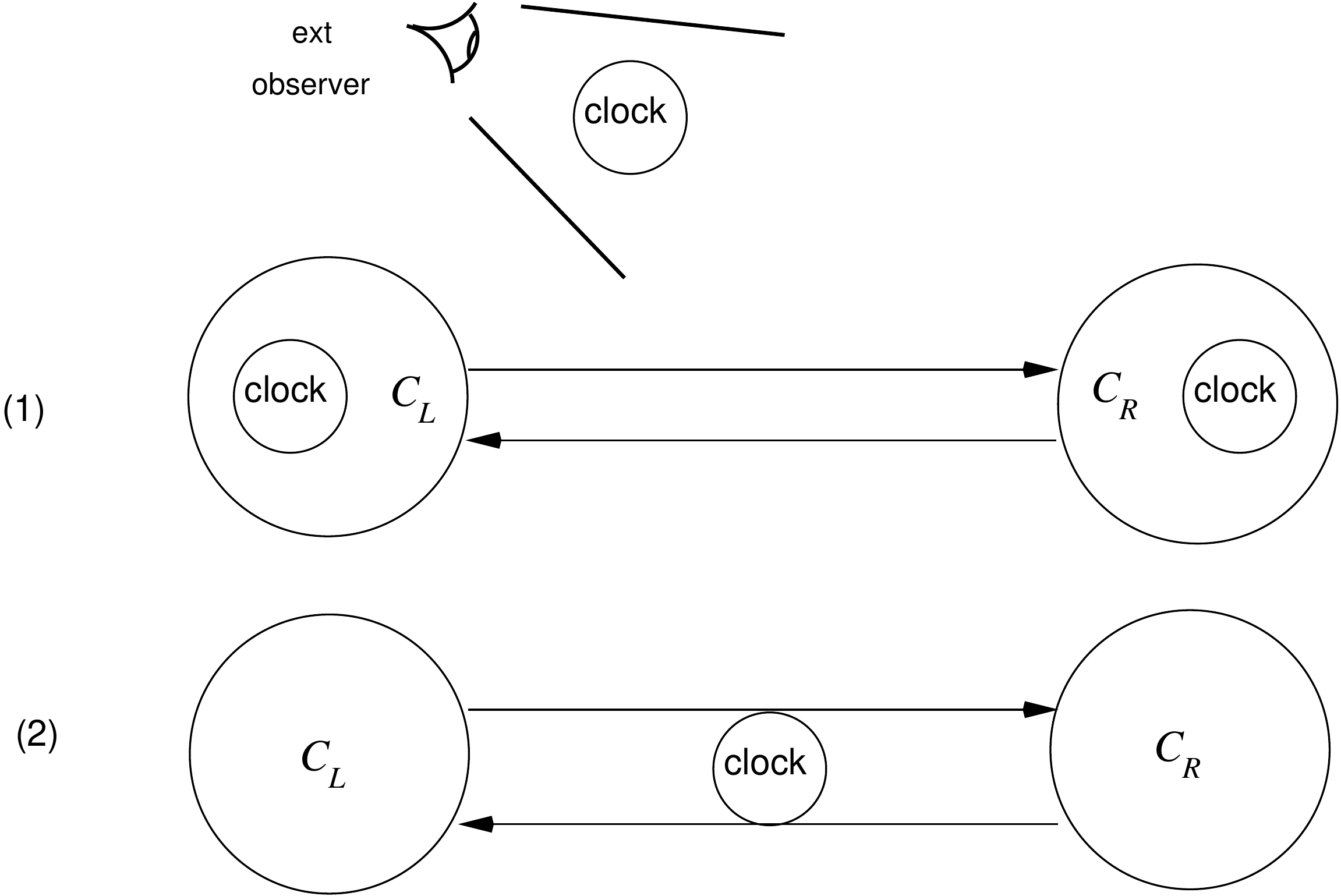}
\caption{\small Two time models: (1) asynchronous and (2) synchronous models
are about where each agent's clock signal is sourced.\label{synchasynch}}
\end{center}
\end{figure}
When agents are independent, they can
each maintain independent state, and hence have independent clocks;
but when agents are entangled, or co-dependent, they share all the
state that pertains to their co-dependent promises, including a
common clock. The two cases are shown in figure \ref{synchasynch}:
\begin{enumerate}
\item {\em Local (weak entanglement)}: each agent can change
  independently and generates its own clock ticks, or its own sense of
  time. Messages may be passed, influencing changes on either side
  with weak coupling, so changes and observations can also be
  interleaved (independently) while waiting for messages to be
  passed.
 Partial or weak  entanglement implies the existence of an independent
  {\em internal event clock} at each cell: the agent can make and act on promises without
  dependency on its counterpart, except when it comes to sending
  \tick-\tock packets.  \beq
  C_L &\promise{X_L}& C_R\\
  C_R &\promise{X_R}& C_L\\
  N_L &\promise{\Ltick \;|\; \Rtick}& N_R\\
  N_R &\promise{\Rtick \;|\; \Ltick}& N_L \eeq The link (as an
  independent sub-entity of the {\em pair} of cells) is constrained to act
  as a single unit, but other aspects of the agents can make unrelated
  promises without depending on the other agent, e.g.  observe the
  link and perform other functions at any time, according to the ticks
  of their independent clocks. For example, they could observe the
  link and detect if it had stalled. This weak coupling is essential
  for scaling beyond more than one interface per cell.

\item {\em Non-local (strong entanglement)}: both agents are dependent
  on the ticks from their mutual interaction, and the messages passed
  between them are the only clock they know. All other changes on
  either side are strongly dependent on the message passing.  Complete
  entanglement implies a {\em shared event clock} for the whole cell:
  the entire cell cannot act or promise anything without being in
  possession of the \tick-\tock token:
\beq
C_L &\promise{X_L | \Rtick}& C_R\\
C_R &\promise{X_R | \Ltick}& C_L\\
N_L &\promise{\Ltick | \Rtick}& N_R\\
N_R &\promise{\Rtick | \Ltick}& N_L
\eeq
These can be compared to (\ref{e1}), (\ref{e2}), and (\ref{e3}).
They describe a synchronous, event driven model in which nothing can happen
until an event is received from the other agent. The clock for both
parties is the link itself.  Network agents have no independent time, which
implies that they cannot make assessments of observations that are not
driven by the link.  With more than two nodes, this becomes immediately
overconstrained and untenable.

Every promise $X$ is conditional on a tick arrival.  The agents could
not observe that the link had stalled because the failure of a tick to
arrive would paralyze them. This makes a complete reliance on strong
coupling risky, because it can experience complete deadlock.

This points towards a split brain model for cells (figure \ref{sw}).
\end{enumerate}
If cells are completely bound by a strong entanglement constraint, they are
entirely hostage to the successful keeping of tick-tick promises, and
cannot observe promise state independently in order to detect the
stalling of the basic tick-tock promises.  If part of a cell is only
weakly entangled, it can observe and assess broken promises
independently. This suggests an internally split brain approach in
which network interfaces (see figure \ref{sw}) are strongly entangled
but other parts of a cell are only weakly entangled.

In effect, an entangled link moderates the flow of information on both
sides by (b)locking observability of state. The challenge in using
this as a technology is to encapsulate the promise to transfer data
such that each packet can only be observed on one side or the other.

The conundrum with this arrangement (see figure \ref{sw}) is that
the passage of time will never be single-valued throughout an application
unless we give up locality. Agents need to have a split brain 
approach to time in order to i) be able to maintain strong entanglement
promises, and ii) to be able to observe when all activity has ceased on
a link, in order to restart it.

\subsection{Knowledge propagation (certainty)}

To understand packet delivery with `knowledge' guarantees,
implementation is based on irreducibility of ticks measured by a shared
clock\footnote{In promise theory, knowledge is defined by statistical
  assessment of the keeping of promises, so this is not knowledge as
  defined in \cite{promisebook}. The need for repetitive confirmation
  may be relaxed in the case of highly constrained contexts, such as
  primitive machinery of the kind we expect in network interfaces, so
  we use the term knowledge loosely to really mean information.}.  The
reliable transmission of information, promised by entangled agents,
may be used to promise a shared state machine, inferring each others'
state transitions based on messages passed, and thus to effectively
`know' certain things about the state of the co-dependent agents.  This
non-local determinism forms the basis of a throttle on data observability.

Promise theory indicates that certainty is built on the trusted cooperation of
individual agents\cite{burgesstrust}.  Entanglement at the level of intentional agents is
fragile to the misbehaviour of agents.  When signals are
sent, each side expects them to be accepted, and acted on in an agreed
manner.  If agents lie to one another, all bets are off.
The basis assumptions are:
\begin{assumption}[Synchronous determinism]\label{headermust}
  In an entangled link, short control messages (headers) will always be accepted
  into a dedicated register, if received by an agent at the end of the
  link. Longer payloads, destined for the applications beyond the
  link, may not be accepted into a buffer. $\Box$
\end{assumption}
This implies that the underlying message passing control channel of the
link cannot be halted, barring some intermediate catastrophe.

Once primed, the entanglement of end points can form the basis of a
simple pendulum/pump/motor, which in turn acts as a clock or generator
for transfers. Once the $LR$ symmetry has been broken by insertion of
a message (see 7.4.1 in \cite{promisebook}), it will be superposed
onto the control channel and passed from one side to the other, {\em
  if and only} if the payload can be accepted by the other side.  This
assumes that:

\begin{assumption}[Agents are reliable and trustworthy]
All promises are mutually kept and agents are
  trustworthy during all spacetime events, such as message arrival and
  transmission. $\Box$
\end{assumption}
It is possible for entanglement to be broken, and each of the cell
endpoints would be aware of there being a broken promise {\em if and only
if} the cells were weakly entangled (in a locally split brain picture).

\subsection{Homogeneity of agents (spacetime)}

\begin{figure}[ht]
\begin{center}
\includegraphics[width=4cm]{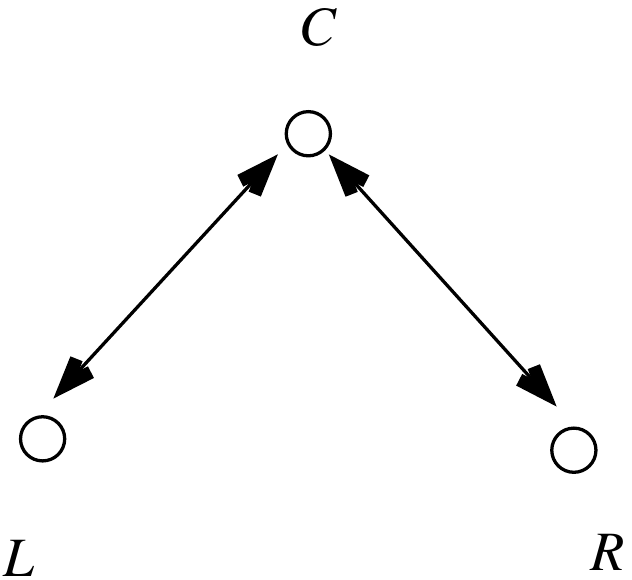}
\caption{\small Collaboration requires a trusted calibration. The
  calibrator could be any implicit `godlike' observer, or permanent
  non-local synchronization, e.g. use of common software. This is
  analogous to having the laws of physics the same on both sides of
  the link.\label{ttp}}
\end{center}
\end{figure}

The ability to trust agents effectively assumes a standard calibration
of both ends of a link against an impartial third party (see figure
\ref{ttp}).  This trusted party might be common software, or a third
party service, but it must exist, else no agent can be sure of what
its neighbour will do with data it attempts to send (it is analogous
to having the same laws of physics at both ends of the
link)\cite{spacetime1,spacetime2}.
\begin{assumption}[Spacetime homogeneity]
  All agents keep the same homogeneous basic set of promises,
  according to their agreed left/right roles, because the
  effectiveness of entanglement promises depends entirely on a uniform
  conditional basis. $\Box$
\end{assumption}
This is essentially an assumption of non-local trust in the basic
behaviours of cells.

\subsection{Message semantics}

The exterior promises made by messages are:

\begin{itemize}
\item All or nothing outcome of data transmission (from $Q_S$ to $Q_R$)
\item If we consider the sending of a packet from the first imposition
in (\ref{init}) as an operator $\hat P$, acting on a data state $|S\rangle$,
then the operator has semantics:
\beq
\hat P | M\rangle = | M+P\rangle 
\eeq
but the interior time it takes is undefined (four or more entangled ticks).
The exterior time is one exterior-tick.
\end{itemize}

\subsection{Transactional semantics, signal heuristics}

A trusted shared-state is maintained by copying local state, from each
side, in a continuous chain, and make second order inferences, which build on the
trust in the behaviour of the endpoints.

Unlike more usual consensus systems, we are building consensus not
about state but about local conservation of data. Data are
distinguishable and countable, and there should be neither loss nor
duplication of data agents.

Given a declaration of primitive state, an agent can claim
instantaneous knowledge about single primitive facts.  This is
not `knowledge' in the full promise theory sense of accumulated
certainty, but more like mutual information. 

The promises are encoded as bits in the headers of packets, for packet
level transfer, and in the message bodies at the message
layer\footnote{A confirmation of the specific message integrity, by
  return of a delivery hash may be embedded in a header as an
  implementation detail.}.  The ability to depend on the entanglement
needs a second order confirmation of receipt:

We provide three equivalent descriptions of the four stages with
different perspectives. The first heuristic schematic involves
four promise steps:
\beq
S &\imposition{~~~\teck(\text{Here's what I know})~~~}& R\nonumber\\
S &\revpromise{\tack(\text{I ack that I know what you sent me})}& R\nonumber\\
S &\promise{\tick:\text{I can depend that you know what I sent}}& R\nonumber\\
S &\revpromise{\tick:\text{I can depend that you depend on me}}& R,\nonumber\\
\label{4ph} 
\eeq 
In the second, we may interpret them as follows:
\begin{enumerate}
\setcounter{enumi}{-1}
\item Recipient: I know nothing, Mr Fawlty.
\item Sender: (\teck) Here is what I know 
\item Recipient: 
\begin{itemize}
\item (\tack) I now know what you know
\item (\nack) Que? (return to 0)
\end{itemize}
\item Sender: (\tick) You received my last message, so I know that you know what I know, provided you haven't forgotten it.
\item Recipient: (\tock) You can depend on me knowing that last thing you said, assuming that I haven't forgotten it.
\end{enumerate}
The second promise is no guarantee that the agent will use data it was
sent. To make this binding, in a verifiable sense, is impossible
within the scope of the link, except perhaps by analysis of long term
repeated misbehaviour at the cell level.  $S$ and $R$ agents can only
`take or leave' what the other offers. The hope is that this
stabilizes into an entanglement on which all other certainties can
be built.

Consider the promises needed to transmit a single packet reliably from
$N_S$ to $N_R$, in such a way that $N_S$ and $N_R$ promise the
location of the packet.  We assume that the buffer queues for the
messages are `externally observable state', while promises made by
$N_i$ are not externally observable; they are only on the interior of
the link.

Readers may wonder if the keeping of a data promise may ultimately be
satisfied by sending several packets, when a recipient fails to
receive a transmission on first try. In other words, is there `retry'?
No promise need be one-to-one correspondence with a packet attempt.
Several packets may be sent to keep a promise, on the interior of the
link, without breaking protocol, and these retries would not be
observable to the cell\footnote{Once we enter the realm of multipath,
  multihop networking, at the cellular level, a more complicated story
  is needed concerning the idempotence of signals (see section
  \ref{idemp}), because assurances about the link cannot replace
  assurances about what is kept in intermediate buffer memory, which
  becomes part of the effective linkage, at the scale of a message.  This
  topic is deferred to a sequel.}.

The third form of the promises in (\ref{4ph}), may now be spelled out 
at a more technical level. We refer readers to appendix \ref{thepromises}
for these details.

\subsection{Assessment for packets}

Have we succeeded in maintaining the integrity of a single packet state?
How many copies of $P$ were observable in the network? The latter is a
slightly tricky question in a distributed system, because of the
subjective experiences of observers, i.e. special relativity. 

Figure \ref{ttp} shows the structure of an observation of the two ends
of a link.  A godlike observer with infinite powers of access might
observe zero, one or two copies of a packet, depending on when
measurements are taken, on its clock. However, no real agent has such
access; each must observe changes available to it, by its own clock.
Within an irreducible superagent, only a single copy of a queued
message $Q(P)$ exists at the buffer queue $Q_R$\footnote{This requires
  some justification, however, since the link alone cannot make this
  promise; idempotence of $P$ must also play a role (see section
  \ref{idemp}).}.  Other copies of the same information, beyond the
link, are naturally outside the scope of discussion.

The interior signalling, within the entangled link, promises exterior
certainty about which side LR/SR of the link a packet can be
observed reliably, just as long as each side keeps its promises deterministically. One
cannot discount the possibility that promises might fail to be kept
for unknown reasons, no matter how isolated and apparently
deterministic the network agents might appear to be.

Whether $S$ can safely delete its copy in steps 2, 3, or 4 is
debatable. The earliest moment at which it could assume that the
message is passed is on receiving an acknowledgment.  There is no
compelling reason to wait for confirmation, except that $R$ wants to
know that $S$ intends to delete its copy, which it would not do if it
did not receive the acknowledgment or if it died in the meanwhile.
There is thus an additional level of certainty in making one more
cycle to add the confirmation, which we shall assume henceforth.

A unique semantic label (like a hash), as part of
$\tack(P)$, would make confirmation more precise.  However, trusting
the behaviours is necessary anyway, so it might be considered
redundant in that respect, if one believes the indeterminism of the
link has been effectively expunged.  A cheaper alternative might 
be sufficient---after all, we have assumed (assumption \ref{headermust})
that headers must always be received and accepted by the network
agents.

Since these interactions are expected (unlike the initial imposition
of a packet, we can probably assume that there are no reasonable
impediments to receipt of a $\tack(P)$, and thus being alive is
sufficient cause to infer that the message was received, and that
agents don't forget what just happened to them. In this approach, agents
promise to give up their autonomy and become entangled intentionally,
and the entanglement is what reminds them of this. 

The final stage is still ambiguous if the alive message does
not come back. Then the link stops altogether, and neither side
notices since their time is driven by the exchanges.  This last matter
is essentially the analogue of the FLP proof that distributed
consensus is impossible in finite time. The workaround here involves
stopping time itself while the job is done, relative to other parts of
a wider network. The cost of certainty is temporary paralysis.

\subsection{Assessment for messages}\label{idemp}

The promise of once-only delivery cannot be trivially extended to
multi-part multi-hop messages, in more complicated topologies, without
some work. We must defer the full discussion for a sequel, and make
only a few remarks here.  It is {\em possible} for multiple copies of
a packet to be observed, duplicated, and transmitted around a network,
if agents fail to keep the necessary promises, no matter whether out
of ignorance or malice. This is not specifically a weakness of our
scheme: it is not easy to promise a negative result.

Nor is it possible to prevent unexpected behaviours: since no agent
can make a promise on behalf of another. There are two pragmatic ways
to localize the responsibilty for intended outcomes to the end points,
away from intermediate interference:
\begin{itemize}
\item One is to used shared secrets or encrypted messenging to make
  corruption by `man in the middle' interference detectable. This need
  not be promised at all layers in a communication stack: high level
  encryption would suffice for detection by the intentional agents.
  See the notes in section \ref{mim}.

\item Another way is for each packet to make a separate and uniquely
  labelled promise (see section 3.12 in \cite{promisebook}) by
  promising a unique desired state.  If each unique and intentionally
  different promise has its own label, and then duplication may be
  detected, assumed redundant, and ignored idempotently.
\end{itemize}
Idempotence of promises means that a promise
repeated $n$ times is the same as the promise given
once\footnote{Telling you twice that we owe you 100 dollars doesn't
  mean that we owe you 200 dollars.}: 
\beq 
\left( S \promise{+M} R \right)^n = S \promise{+M} R.  
\eeq
Idempotence must play a role in promising uniqueness, where we don't
have complete control over causation.  Just as endless \tick and \tock
cycles promise nothing new (except freshness), so repeated intentional
messages signal nothing new, and no advancement of state. Efficiency
can, of course, be compromised by excess copies (the cost of
repetition), but no confusion or duplication of intent would be
signalled, provided distinguishability were managed properly, by
idempotence of promises. The bottom-up design of the entanglement networks
seeks to minimize this possibility of duplication, but we have to defer that
discussion to the sequel.

\subsection{Man in the middle: interference and intentional forwarding}\label{mim}

What if an agent could insert itself into the middle of a link and
imitate the end points $C_L$ and $C_R$? Would this invalidate the
promises of entanglement, as in the quantum mechanical case?  This is
quite easy to do, in principle, because no secret knowledge is
required to run the protocol. Such an insertion of an intermediary
could be used as a feature or as a bug (an attack). The insertion of a
switch or router for multi-hop forwarding uses precisely this approach
to deliver packets by chains of voluntary cooperation (see chapter 11
of \cite{promisebook}). Alternatively, a wiretap insertion for
breaking security promises would be considered an intrusion.

From promise theory we know that the insertion of intermediate agents
renders unconditional promises impossible, because of the basic locality
of promises---that no agent can make promises on behalf of any
agent other.  To establish a similar level of assurance to
keep promises through third parties, one acquires the burden of a web
of conditional assurances, which grows like the square of the number
of intermediaries, unless complete trust in intermediate agents can be
assumed. This is similar to the design and cost of building
blockchains (the principle is the same). 
This is a complicated topic, so we shall only make some simple
points here, and defer a proper discussion for a sequel.

Between a sender $S$ and receiver $R$, an intermediate agent $I$ could
receive packets and pass them on without alteration, invisibly tapping
the channel. Or it could become a new endpoint, blocking transmission
in one direction and masquerading as the blocked agent. In either case
the point to point protocol cannot protect against such an abuse of
intent, so long as there is no authentication of the agents.

\begin{figure}[ht]
\begin{center}
\includegraphics[width=4cm]{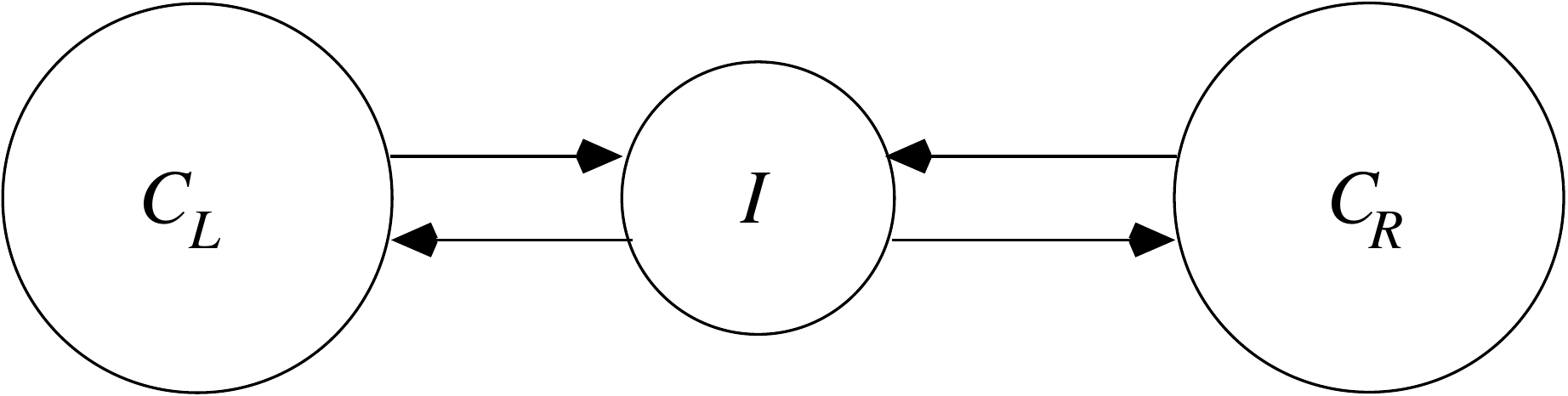}
\caption{\small What if an agent could insert itself into the middle
  of a link and imitate the end points.  This is quite easy to do, if
  no secret knowledge is required to play the protocol
  game.\label{link2}}
\end{center}
\end{figure}
Could the end points detect tampering without extra bits? At the
physical level, this is doubtful.  In the current implementation, any
agent can act as an endpoint, and there is nothing to distinguish any
agent from the next.  What we understand, implicitly from blockchain\cite{bitcoin,blockchain,bcsummary1},
is that detection of tampering requires something like the
longitudinal entanglement of all agents in a chain, or a binding to a
trusted third party. The problem for a network protocol is that
explicit trust is broken by every inserted host in a chain. Network
agents cannot forward data without going through a external host node,
so to extend the current approach to multi-hop architectures requires
full trust in the entire infrastructure, as well as additional
reasoning.

Ignoring the multihop issue for a moment, a single wiretap would be
enough to lead to spoofing. An intermediate agent would be able to
fool sender or receiver into believing in is a consensus when there
were, in fact, none. These concerns can be addressed in various ways,
from physical isolation to encoding measures.  At the packet level,
one can only promise to know that the last tick or message was
received, not whether a subsequent acknowledgment was sent, but not
received. Links can therefore be stalled and spoofed here too.  A
responsibility for the integrity of knowledge, like other properties,
is pushed to the ends of the link. One way to build in tamper-proofing would
to entangle sequential message packets by a simple encryption of the packets,
essentially by cipher blockchaining.

The lowest level physical links are those most vulnerable to the
physical security of their wires and channels.  Higher level
derivative entanglement could more easily embed privacy through
encryption, making detection of tampering straightforward.  These
matters are subtle and we shall not discuss them further here.

\section{Entangled link functionality}

The purpose of an entangled link is to encapsulate deterministic
transmission of data, transparently, packet by packet. By
equilibrating a message $M_S$ to an identical image $M_R$ at another
location, before revealing it to the recipient, we can increase the
knowledge about consistency of state for mission critical applications. This
method is not a panacea, and must work in a suitably curated context,
but we contend that it offers a foundation for reliability.

Philosophically, the idea of entanglement opens for a discussion of
many deep ideas about relativity and mutual knowledge, and how the
concept of `consensus' can even make sense across a spatial region,
limited by the propagation of packets. These questions are
familiar from modern physics, but their information theoretical
counterparts are only just being appreciated. It is not the place
for that discussion here, but informed readers may recognize 
the issues that connect spacetime and scale to the propagation of
information.

Our motivation is that we might use this bottom-up approach to build
distributed applications, in which key data are in a sychronized state
at all times. This seems plausible, either within a datacentre, or
even across the planet, with certain provisos. The essence of a
solution is how to stop the clocks for certain observers, over a
bounded spacetime region, while hidden processes continue to work
unnoticed within. It is a quantization of observability over a spacetime interval.

This is not a new idea: current approaches to consistency using
locking\cite{chubby}, master-slave systems\cite{paxos,raft}, or
blockchain\cite{bitcoin,blockchain,bcsummary1}, to similar effect,
operating across TCP/IP.  All scale with significant costs.  In our
approach, we use several signals back and forth between each node for
every packet.  What we win from this is sychronized data at every
observable step. One has to be cautious about the scaling of promises
claimed, as entanglement and consensus may be no less expensive to
maintain, but the reorganization inspired by information theoretical
entanglement could help to optimize the problem.
Centralized coordination, either by locking,
queueing, or calibration services, atop TCP/IP, works well enough for
many cases where one can solve consistency by brute force. Our
approach lends itself to a different kind of lightweight transactional
network architecture, designed for determinism in each step. The virtues
of our approach remains to be discussed in detail.

In the remainder of this section, we review the core concepts, which perhaps become
buried in the technicalities, with a more pedagogical eye. 

\subsection{Agreement}

Consensus (multiparty agreement) arises between two agents when they
agree (see section 8.4 in \cite{promisebook}), The steps for promise
theoretic agreement are:
\begin{enumerate}
\item $(+)$ Share an invariant promise proposal.
\item $(-)$ Agent 1 observes/accepts proposal.
\item $(-)$ Agent 2 observes/accepts proposal.
\item $(+|-)$ Agent 1 signs proposal if accepted.
\item $(+|-)$ Agent 2 signs proposal if accepted.
\item $(-)$ Agent 1 observes signature.
\item $(-)$ Agent 2 observes signature.
\end{enumerate}
There is an implicit partial ordering in these steps, encoded via
conditional promises, and their implicit order.  After all promises
have been kept, both agents can be said to have agreed or reached a
consensus, and a state of common knowledge.

The steps can be simplified, when the proposal comes from one of two agents:
\begin{center}
\small
\begin{tabular}{|c|c|l|}
\hline
& \sc Agent & \sc Promise/Intent\\
\hline
1 &$S$& Share presigned proposal\\
\hline
2 &$R$& Accept proposal and signature\\
 && and ack. by signing proposal\\
\hline
3 &$S$& Accept $R$'s signature\\
&&($S$ and $R$ now agree and know it)\\
\hline
4 &$R$& Receive accepted signature from $S$\\
&&(everyone finished)\\
\hline
\end{tabular}
\normalsize
\end{center}
This is the version we use for running a transaction protocol.
The steps are encoded into the protocol for entanglement
as follows:
\begin{center}
\small
\begin{tabular}{|c|c|c|l|}
\hline
& \sc Agent & \sc Signal & \sc Promise/Intent\\
\hline
1& $S$& \teck & Share $P$ into $N_S$\\
\hline
2& $R$& \tack & Copy $P$ into $N_R$\\
&     &       & send acceptance\\
\hline
3& $S$& \tick & Delete $P$ from $N_S$ and\\
&     &       & (set $P$ in $Q_S$ not observable)\\
\hline
4& $R$& \tock & Move $P$ from $N_R$ to $Q_R$\\
&     &       & (make $P$ in $Q_R$ observable)\\
\hline
\end{tabular}
\normalsize
\end{center}

\subsection{Timescales}

There are hidden assumptions behind promises of consistency.  The
first crucial assumption is that the proposal or desired outcome is
invariant over the lifetime of the consensus process, else one could
not stabilize a transmission in a particular direction.  The proposal
exposed in the first step of the agreement process, described in the
previous section, may not change as the agents go about their interior
promises to observe, copy, and agree to it. In the language of clocks,
the object of agreement, in any common knowledge problem, needs to be
persistent on a timescale longer than the promises to abide by the
intermediary steps.  So, whereas one typically talks about
`correctness' in computer science (a semantic assessment), it is more
a question of stability (a dynamical assessment), or invariance of
assumed targets, during the key change
processes\cite{certainty,promisebook}.

So, to converge on a stable target, it is assumed that the
timescales (as measured on the clock of some godlike observer) for the
lifetime of the exterior promises to transmit with integrity, must be
significantly longer than the timescale over which the interior
promises are defined and kept, by a good margin, else one is racing
against a moving target: \beq \Delta t_\text{exterior} \gg \Delta
t_\text{interior}.  \eeq It is not coincidence that this is also the
condition of equilibrium, with equilibration or `relaxation' time
$t_\text{relax}$ for transactions: \beq \Delta t_\text{exterior} \gg
\Delta t_\text{relax}\gg \Delta t_\text{interior}, \eeq or
equivalently: \beq \Delta t_\text{common knowledge} \gg \Delta
t_\text{transaction}\gg \Delta t_\text{\tick}, \eeq In our technical
implementation, this translates into the rates of interior \tick/\tock
processes relative to the rate of new messages $M$.  \beq \Delta t_M
\gg \Delta t_P\gg \Delta t_\text{\tick}, \eeq We expect to be able to
achieve consensus about $M$ is there are many more \tick events than
there are new messages. This (hopefully) seems like an obvious point,
but most consensus discussions brush over such limitations. The upshot
is that an ideal technology should seek to maximize the rate of
interior equilibration. Faster is always better.

\begin{figure}[ht]
\begin{center}
\includegraphics[width=8cm]{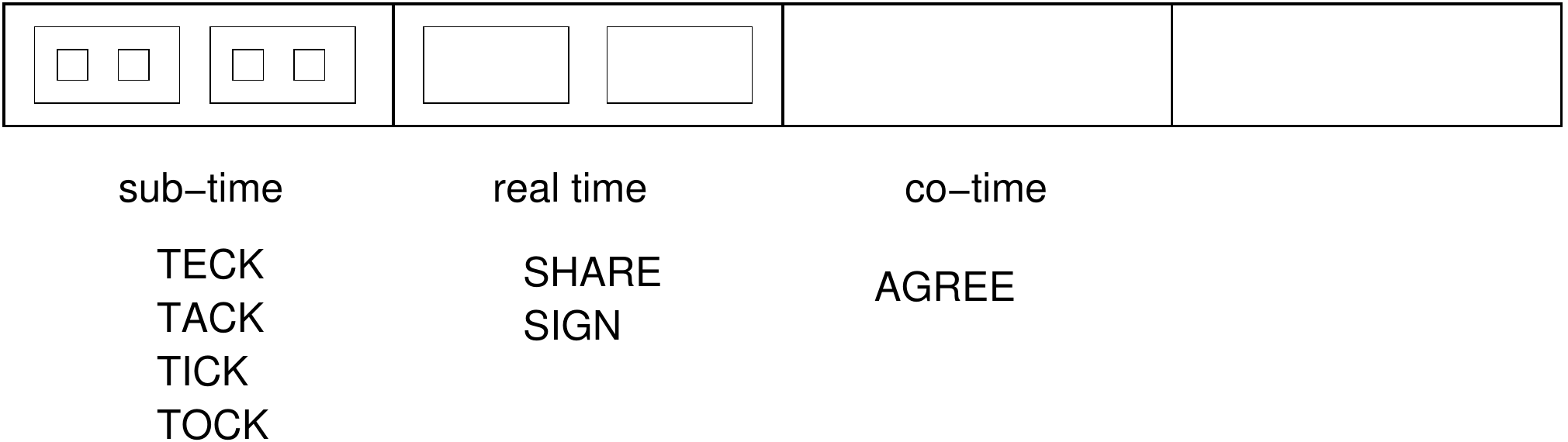}
\caption{\small Locally, the composition of timescales for
  quantization of certainty may be viewed at three scales: individual
  subtime transactions, roundtrip times, and co-time assured
  transfers. Using the analogy of version control, only the co-time
  units are `committed' as accepted new versions of the current
  transaction.  Equilibration time is kept comparable to subtime,
  according the the observer's clock at each end.\label{subtime}}
\end{center}
\end{figure}

The core assumption (often unstated) is, therefore, that consensus
equilibrium systems is that the data cannot be allowed to change
faster than the superagent can reach equilibrium or consensus.
Moreover, things that depend on the value need to be frozen for the
duration of the message transfer, so $M$ is a slowly varying quantity.

\subsection{Playing with time in a split brain world}

The split brain model, within a cell, allows a cell to detect when a
link has stalled. To accomplish this, each cell maintains effectively
two kinds of clocks: one engaged in link activity, and one sampling
the other for stalled state. The clock that drives each link, shared
by the endpoints of a link, is thus watched over (on each side) over
by processes synchronized by the cells' clocks, which can observe all
the network interfaces they are connected to.  If message exchanges
time out, according to the observer clock, some recovery is in order.
Recovery actions depend on the larger topology, so we shall not
comment on details here.  Alternatively, the link simply dies, and
time effectively stops for the entangled parties. No data are sent or
received. It may be possible to restart the stalled link in some cases
(by repairing a broken wire, for example).

\subsection{Scaling consistency}

A collection of consistent transactions must lead to a consistent
collection, regardless of how we packetize messages. However, it is
not obvious that other forms of composition, such as end-to-end serial
compositions of links, can automatically assume the same promise of
consistency (see chapter 11 in \cite{promisebook}).  Promises are not
transitive properties, and do not therefore extend across intermediate
agents, without more effort. This means we can't automatically assume
entanglement properties of a journey composed to several legs.

Because of the timescale constraints, entanglement works most effectively
over `small' amounts of data, and small regions of spacetime, but
becomes untenable as we try to include more information, because the
ratio of interior to exterior promises grows, creating long
relaxation times.  This is the same as for classical consensus protocols,
but the difference lies in the fine-grained observability of the process,
which opens for new routes to certainty and recovery in case of trouble.

Network links only have meaning in the context of a larger process,
some of which extends beyond the region of entanglement.  If an entire
cell crashes on one end of a link, and loses a large part of its
partially sent or received message queue, we cannot say anything about
recoverability of the state of the cell in the future, or what the
link's recovery might do to it.  A consequence of entangling the link
is that there is an implicit coordination of the connected cells too.
In order to recover one cell's state, neighbouring cells' states may
play a part in the recovery.  Catastrophes at a scale of cells 
provide no automatic context for an automated recovery. Application
recovery is therefore a separate issue.  This is a normal scenario, in
any technology, and it means that entanglement and recovery can only
work within reasonable bounds.

Once a data packet passes outside a region of entanglement (the
interior of a link), its value can drift away from that on the other
side of the link, independent and unconstrained.  Because equilibrium
takes (interior) time to establish, a unified exterior state can be
promised only after these interior promises have been kept. The key to
quickly promising consensus lies in managing the scope (or spacetime
region) over which data are equilibrated.

Within the bounds of an entangled link, this is handled by quantizing
`events' in such a way that intermediate states are not observable.  
One can try to scale this approach, for higher level communications too, 
by defining a new meta-process on top of the lower level
processes, that spans multiple entangled links, and passes a new level
of \tick/\tock messages end to end. The whole story may thus be
repeated, at a slower rate and on a wider range, forming a reliable
tunnel, analogous to figure \ref{sw}:
\begin{itemize}
\item Forming a split brain within the application, and equilibrating
  the state of {\em application} endpoints $A_S^{(+)}$ and
  $A_R^{(-)}$, application packet by application packet, over a
  distributed tunnel.

\item Using destructive observation to teleport packets from application $Q_S$ to
  application $Q_R$.  
\end{itemize}
Entanglement turns a passive equilibrium into an active co-determined
one. A change of state on either side propagates instantaneously and
deterministically in exterior time to the other end of the link,
just as in other distributed consensus databases.

\subsection{Impossibility theorems}

The well known FLP proof and the `General's Problem' about consistency
and asynchronous uncertainty challenge the designs for approximating
consensus in contemporary technologies. Solutions building on TCP/IP
networks tend to focus on the state of data, rather than on the
timescales over which the states can be promised, since delivery times
are not easily quantifiable in TCP/IP networks. Our observation here
is that time plays a crucial role in the meaning of determinism, and
that one can use this to greater advantage in a reprioritized implementation.

The FLP result says that consensus is impossible, in an asynchronous
system, where agents only {\em might} be unreliable. The simplest
improvement might therefore be to avoid the indeterminism of asynchrony, to the extent
that this is possible.  The strategy in an entangled link is to
engineer synchronicity back into the communications by adding a
management layer `underneath' conventional communications (or in
promise language, on the interior of the network agents).
Asynchronous behaviours get coarse-grained away by restricted
observability, leading to an exterior promise of quasi-synchronous
transmission.  This is a workaround of the FLP result.

\subsection{Failure modes}

If a message does not arrive, for some reason (there are not many
plausible reasons for failure at the level of entanglement), then the
link itself may simply stop transmitting. With a split brain
cell agent, this is detectable, even as the link dynamics
are given the primacy to drive progress. Comparing to a TCP/IP delivery:

\begin{center}
\small
\begin{tabular}{|l|l|l|}
\hline
\sc Condition & \sc TCP/IP & \sc Entangled\\
\hline
Send & Impose / Collide & Scheduled slot\\
\hline
Not ready & Drop packet, recover & Try again.\\
\hline
Queue full & Drop packet, recover & N/A\\
\hline
\end{tabular}
\normalsize
\end{center}
The additional certainty of message delivery means that the need for
an infinite number of messages (the Generals' Problem) is formally
rescinded. With fully independent clocks, endpoints might assess a
failure in the {\em imposition} of an asynchronous message at an
inconvenient moment. In an entangled link, messages are {\em promised}
according to an agreed schedule, and this schedule literally stops time for
both parties until recovery is possible.

This implicit notion of time is fundamental to the process of
acknowledgement.  Normally one does not take into account the
relativity of the agents; in our case, relativity is built into the
design.

\section{Building recursive entanglement}

We are focusing here on the engineering principles involved in keeping
single link promises, both theoretical and practical.  The key approach
is to scale the definitions of agents and packets, such that the
indeterminism of the interior agents promises are reframed as
quasi-deterministic exterior promises.  The next steps are to build
assurances over wider areas, spanning multiple hops. This introduces
plenty of new issues to be discussed in a sequel.

The entanglement method is extensible to arbitrary levels of
abstraction, in principle.  Being able to promise reliable point to
point delivery allows one to reason about delivery at any scale, but
it does not ensure the inevitable correctness of messages, which are
the responsibility of applications. This requires a trusted platform
too.

Given the cost of entangling agents, including the energy cost
of maintaining a \tick/\tock, this might not be a method one would suggest
lightly for communications in any context; however, in more stringent
circumstances, it seems well suited for assuring data
replication, e.g. in mission critical scenarios, or disaster recovery of
high value data.  Further issues have to be discussed before we can
build applications on top of it.

This includes the routing of messages through a full network, and also
the effects of serial composition of links, each of which
independently plays with time.  The rapid circulation of tokens could
make multi-hop journies more sensitive to channel differences. We
might foresee the possibility of `timing storms', or unstable modes of
oscillation, at the interfaces between hops with different relative
rates of promise keeping. This could require some damping mechanisms
to be incorporated, over wider areas, especially on inhomogeneous
networks with uneven latencies.

In short, to scale the entanglement, we may not automatically assume that a
serial composition of links will keep the same promises as
a single link. However, we can reimplement the protocol recursively
on top of the links.
Schematically, the approach will be the same at all scales; one sends a 
promise proposal:
\beq M &=&
\left(
\begin{array}{c}
\text{Intent}\\
\text{Ack of Intent}
\end{array}
\right)
\eeq
at some scale, expecting a complementary form in return:
\beq
\overline M &=& 
\left( 
\begin{array}{c}
\text{Intent to ack}\\
\text{Ack of Ack}
\end{array}
\right) \eeq If one can avoid multiple (redundant) causal pathways,
this interaction remains `simple', else it might lead to new forms of
interference. One way to avoid it is to employ a spanning tree
approach to wide area coverage\cite{borrilltree}.

Each higher level of data payload, in a message $M$ relies on the
promises below it for its atomicity and integrity (as represented by
figure \ref{entangled}).  By reprioritizing network functionalities,
redesigning the layers starting from the bottom, it should not be
necessary to build many layers. The assumption of lower level reliability at
every scale, which may be assured in turn by the application of such
`turtles' all the way down.

\section{Summary}

We have outlined the description and scaling of reliable
point-to-point communications links, based on promise entanglement. We
showed that a system using such links behaves as a quasi-deterministic
system (i.e. one in which cause and effect appear synchronous
according to a shared clock). The entanglement property may be
effective provided all network agents maintain the same uniform set of
promises, and the amount of transmitted information per packet is
small.  This allows fast equilibration of state, and straightforward
reasoning about data delivery, including data consistency.

Once data emerge from the entangled link region, no further promises
about data can be made; e.g. if an entire cell or application crashed,
and lost its runtime state, say, half a transmitted message, we might
not be able to say anything about the relative states of sender and
receiver when it came back. So an entangled link cannot be expected to
recover a session whose larger context has been lost.  This is as
expected, and the connected parties would still have to detect the
collapse, and determine their own response to the failure at their own
level. Such events are likely rare, however, compared to smaller contentions
over network delivery, especially in highly utilized environments like
cloud computing infrastructure; so, it seems likely that there is a
beneficial use case for such an approach.

In the sequel, we shall address the routing of messages between higher
level applications, through multiple ground level cells, in order to promise average
deterministic outcomes from non-deterministic multi-path networks.

\bibliographystyle{unsrt}
\bibliography{spacetime,money} 

\appendix

This appendix describes implementation details of promises and agents
involved in the chain of custody in an entangled link.  There might be
several equivalent representations of these promises, so we seek a
necessary and sufficient fundamental representation.

\subsection{Buffer queues and observability}\label{qq}

For completeness, and precision, we declare the semantics of queues.
Referring to the figure \ref{entangled}, in which cell $C_L$ has the
sender role $S$, with an interior message process $M_S$, and $C_R$ is
the receiver cell with role $R$, and message recipient $M_R$.
\begin{itemize}
\item Data in a packet $P$ begin in the messaging process $M_S$, contained by
  $C_S$, and are pushed into a buffer to send to another agent $C_R$,
which promises to accept without impediment:
\beq 
  M_S &\imposition{+P}& Q_S\\
  Q_S &\promise{-P}& M_S.
\eeq
The information in $P$ does not disappear from $M_S$, but (to account for
its transmission) we enter it into a queue (Last In First Out) structure:.
The contents of this queue $Q_S$ are observable to the message originator $M_S$, including
its network interface transmission register $N_S^{(+)}$ on a packet by packet basis:
\beq
Q_S &\promise{+Q_S(P)}& M_S\\
Q_S &\promise{+Q_S(P)}& N_S^{(+)}
\eeq
The $Q(P)$ notation is shorthand\cite{promisebook} for the queue's representation of the
packet, dependent on its value $P$. This is conditional on receiving $P$, i.e.:
\beq
Q_S \promise{+Q_S(P)} N_S^{(+)} \equiv 
\left\lbrace\begin{array}{c}
Q_S \promise{+(P \in Q_S)|P} N_S^{(+)}\label{e1}\\
 N_S^{(+)} \promise{-P} Q_S
\end{array}\right.
\eeq
We write the normal shorthand $Q_S(P)$ for $Q_S$'s interior representation of $P$, and
$Q_R(P)$ as $Q_R$'s representation of $P$, and so forth, where it is understood that $Q(P)$
has the value $P$.
\item When ready to send, the network interface pulls promised packets from the send queue,
and promises its own representation $Q(P)=P$, based on $Q_S(P)=P$:
\beq
 N_S^{(+)} &\promise{-Q_S(P)}& Q_S,\\
 N_S^{(+)} &\promise{+Q(P) | Q_S(P)}& N_R^{(-)},
\eeq
This is the prerequisite that initiates transmission in (\ref{rts}).
Note that the link representation $Q(P)\equiv\teck(P)$ in our protocol notation.

\item We define $P$ to be `observable at $X$' when
\beq
Q_X \promise{+Q_X(P)} M_X,
\eeq
in other words the message can be received when this
promise is made, and the message is received when
\beq
M_X \promise{-Q_X(P)} Q_X,
\eeq
has been kept.

\item Observability may be rescinded if the promise is deleted:
\beq
Q_S \promise{\neg Q_S(P) | \text{cycle complete}} M_S,
\eeq
in other words, the queue ceases to promise the packet $P$ in its
buffer. When the four phase cycle is complete, the packet $P$
becomes observable at $Q_R$ and non-observable at $Q_S$.
\end{itemize}
The causal determinism rests on making these transitions conditional on the
appropriate set of prerequisite conditions.

\subsection{Interior packet promises, signal details}\label{thepromises}

The four stages of an interaction may be described as follows (see
figure \ref{entangled}).  Readers are reminded that every symbol
representing a promise agent implies independent behaviour, and that
information is only observable by those to whom it is promised.  The
message queues of the sender and receiver play an important role, in
these steps, as the intermediary that decides where packets are
allowed to be seen.  The queue (see appendix \ref{qq}) is the
interface between the split brain entanglement zones of the cells.

\begin{enumerate}
\item The sender $S$ pushes data $P$ into its send queue $Q_S$ (see
  figure \ref{entangled}), which accepts and promises that the
  value is `observable' at the sender queue location:
\beq
S  &\imposition{+P}& Q_S\label{init}\\
Q_S &\promise{-P}& S\\
Q_S &\promise{+Q(P)|P}& S, N_S^{(+)}\label{e2}
\eeq
So $P$ is now observable at $S$ by whatever exterior parties
might try to look (and are promised access).
When $N_S^{(+)}$ accepts it from the queue for transmission,
it promises to share it with $N_R^{(-)}$, conditionally:
\beq
N_S^{(+)} &\promise{-Q(P)}& Q_S\label{rts}\\
N_S^{(+)} &\promise{+\teck(P) \;|\;Q_S(P)}& N_R^{(-)}
\eeq
The registers $N_S$ and $N_R$ are not publicly observable by $C_S$ or
$C_R$, so no observer can see intermediate states of transmission,
until $N_S$ and $N_R$ promise explicitly to update the observability
status of data in the respective queues $Q_S$ and $Q_R$.

The queue buffer is quite important in bounding the scope and
limits of entanglement, i.e. simultaneous co-determination of state,
and according to whose clock. If the entangled state extended beyond, the
cost of entanglement might not be achievable relative to the rate of 
independent changes at either end.

\item The recipient $R$ promises to always accept the \teck packet, but
may promise to accept the \teck packet payload, or not:
\begin{itemize}
\item {\bf Accept}: If $N_R$ accepts the payload $P$,
\beq
N_R^{(-)} &\promise{-\teck(P)}& N_S^{(+)}
\eeq
 then it can promise that it shares the state of the sender,
and it pushes this shared state to the queue $Q_R$:
\beq
N_R^{(-)} &\imposition{+Q_R(P)\;|\;\teck(P)}& Q_R\nonumber\\
\eeq
$N_R$ replies with a \tack to acknowledge receipt (or possibly a \nack to reject it):
\beq
N_R^{(+)} &\promise{+\tack(P)|\teck(P)}& N_S^{(-)}\nonumber\\
\eeq
Now both queues can promise co-determined $P$, but only one side promises to expose it to the
wider cell $C_R$.  This restriction of observability prevents the application clock from
advancing for the world beyond the interface.  The data $Q_S(P)=P$ and
$Q_R(P)=P$ are now duplicates (co-determined by the sub-time exchange), 
and there is overlapping mutual information $P$.
\item {\bf Reject}: If $N_R$ rejects the packet, it cannot claim to share the same information
as the sender, so it does not promise to push anything to $Q_R$, and instead returns a 
\nack message\footnote{In earlier documents, it returned \tock to reject.}:
\beq
N_R^{(+)} &\promise{+\nack(P)|\teck(P)}& N_S^{(-)}\nonumber\\
\eeq
This new state at $R$ co-determines that $P$ should be removed from $N_S$,
and the sequence ends.

\end{itemize}

\item If a packet was accepted, then the significance of the next
exchanges continues to have meaning within the atomic transaction,
otherwise the link does not advance.

{\bf Accept}: $N_R$ acknowledged receipt of $P$ with \tack, 
and $N_S$ promises to accept acknowledgments of receipt, 
\beq
N_S^{(-)} &\promise{-\tack(P)}& N_R^{(+)}\\
N_S^{(+)} &\promise{+\tick\;|\;\tack(P)}& N_R^{(-)}.
\eeq
So $N_S$ infers that its promise to send has been kept, and it
can therefore withdraw that promise without impediment:
\beq
&& N_S^{(-)} \promise{-\tock} N_R^{(+)} \\
&&\left.\begin{array}{c}
N_S \imposition{\neg Q_S(P) \;|\; \chi~} Q_S\\
Q_S \promise{- (\neg Q_S(P))} N_S \\
Q_S \promise{\neg Q_S(P) \;|\; \chi~} M_S
\end{array}\right\}  ~~\text{done~~~~~~}
\eeq
$N_S$ now knows that its promise to send succeeded. However, $N_R$ does not yet
know of $N_S$'s intent to assume completion.

{\bf Reject}: In the case of a rejection \nack, the two equivalent promises are kept
\beq
N_S^{(-)} &\promise{-\nack(P)}& N_R^{(+)}\\
N_S^{(+)} &\promise{+\teck(P)\;|\;\nack(P)}& N_R^{(-)}.
\eeq
and the sequence goes back a step, without removing or altering the promise to
send the packet by $N_S^{(+)}$. The promise to send remains
unkept, and ready to be retried next time it is $N_S$'s turn to
\tick.

\item In the case of successful acknowledgement, a \tick instead of a \teck retry.
We may assume that $R$ accepts a simple \tick, and the queue $Q_R$ 
conditionally promises to reveal its waiting copy of $P$ to $C_R$:
\beq
N_R^{(-)} &\promise{-\tick(P)}& N_S^{(+)}, Q_R\nonumber\\\\
Q_R &\promise{+P\;|\;\teck(P)\AND\tick}& R\label{e3}
\eeq
Now $Q_R$ knows that the receipt is known to the sender, and that it will
not try to resend it, so it is safe for the receiver to reveal its catch.
$R$ signs off by signalling back \tock to $S$, indicating that its copy has been revealed:
\beq
N_R^{(+)} &\promise{+\tock\;|\;\tick}& N_S^{(-)}
\eeq
When $S$ accepts the \tock, it is safe to inform the message $M_S$
application that delivery of $P$ is complete,,e.g.
\beq
Q_S \promise{\neg P} M_S.
\eeq
\end{enumerate}
This completes the detailed autonomous semantics of the signalling of
a single packet $P$ between two agents that are entangled.  Readers
might be surprised at the number of detailed chain of co-dependent
promises that are needed when there is no automatic assurance of
determinism, but this is just accounting\footnote{In fact, the
  situation is not dissimilar to a blockchain, where consensus
  requires longitudinal entanglement of messages, and semantic
  selection of the correct outcome. The difference is that one has no
  way to block clients from seeing transactions that are in possibly
  intermediate stages of completion, so history may need to be rolled
  back with some embarrassment.}.

This sequence differs from a regular TCP like exchange in that
it is effectively synchronous, and a kernel does not have to deal with
duplicates. The split brain may still choose to abort the resending of
a packet after a certain number of retries.

\subsection{Packet agent structure}

A message $M$ is sent by one network interface agent to another. The sender (or
initiator) carries the intent to transmit a message (which is a + promise) with
header $H$, while the recipient is the target of the message. The
recipient responds with a complementary message $\overline M$, whose
header has the complementary structure (a - acceptance promise) with
header $\overline H$.
\begin{figure}[ht]
\begin{center}
\includegraphics[width=6cm]{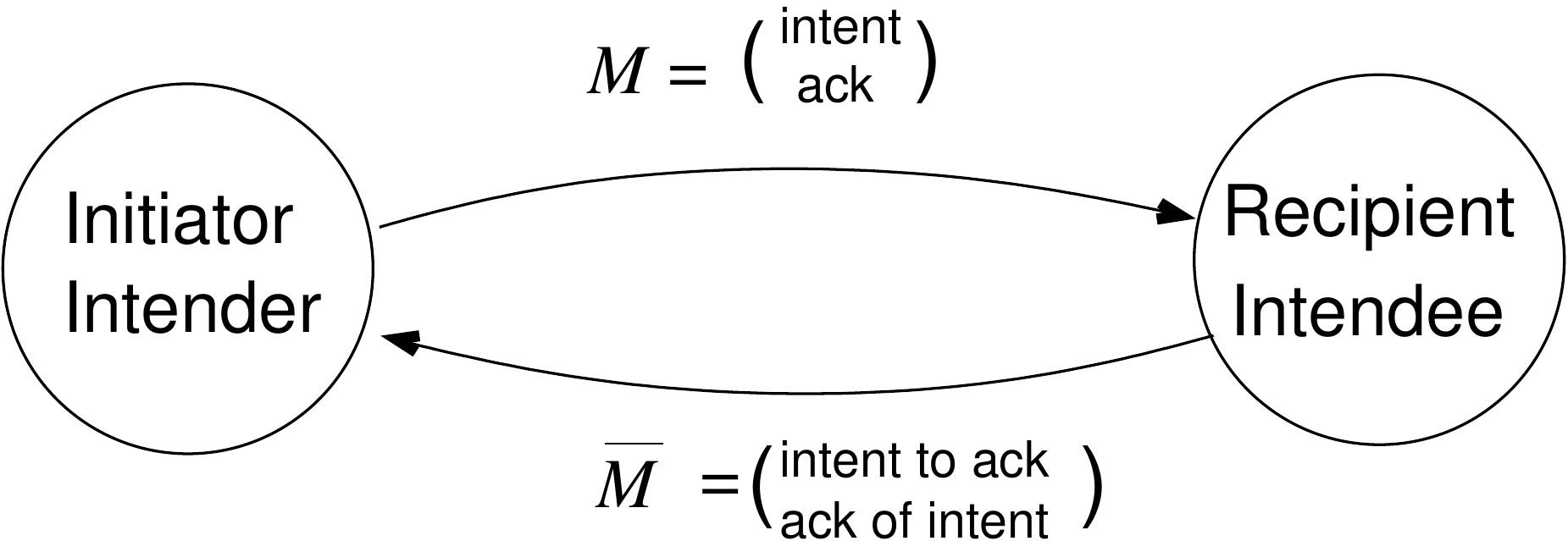}
\caption{\small A message must satisfy this complementary property at
  all times, at all scales involved in message passing. At the link
  layer, intent is kept deliberately simple: to reach the other end,
  and acknowledgment is a simple matter of reflection.  At the level
  of a message, intent involves the arrival of a specific string, so
  acknowledgment is more complicated.\label{message}}
\end{center}
\end{figure}

A message $M = (H,D)$ is a doublet consists of a header $H$ and a
payload $D$. A payload may be optionally empty, i.e. $D = \emptyset$.

\begin{figure}[ht]
\begin{center}
\includegraphics[width=6.5cm]{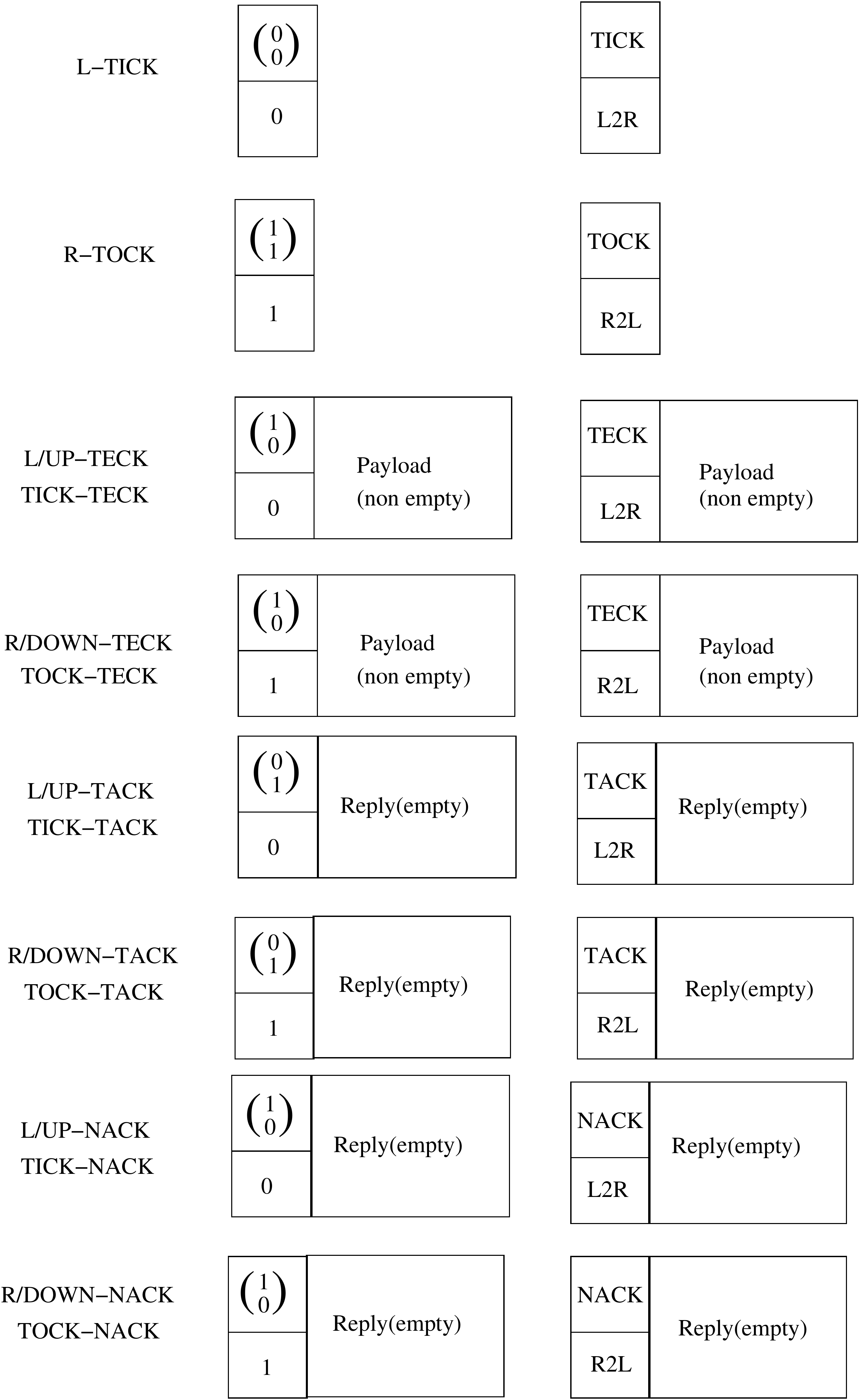}
\caption{\small Message geometries - only \teck messages carry payloads, and there
are two oriented forms of \teck depending on the direction: $\teck_L$ = (tilde) $\widetilde {\teck_R}$.
In this example, we assume that the `left hand' end of the link is the one to send
\tick. A \nack message, which is a negative acknowledgment or rejection
`not ready to accept payload' or `not ack', has the same encoding as \teck, but
with no payload, so there is no ambiguity in coding
as long as one observes the natural condition that \teck with empty 
payload has a special meaning and cannot be accepted
as part of a normal message stream.
Note that the first two are completely anti-symmetrical in the intent-acknowledgment sub-vectors, without any memory or state, and 
thus have no temporal ordering. The subsequent messages are directed from left to right or vice-versa,
indicated by the asymmetry of the intent-acknowledgment parts, where the direction is indicated by the lower 
acknowledgment components; this is the continuation of the \tick-\tock
directional acknowledgment. The intent to send or to reply is indicated by the upper intentional part.
\label{M}}
\end{center}
\end{figure}

\subsection{Header structure}\label{implemH}

Labels $L$ and $R$ stand for `left' and `right', and refer to the
distinct ends of a network connection.  The assignment of left and
right is arbitrary (a result of some dynamical symmetry breaking
process), but we assume it remains invariant during the interactions.
Left is defined as the agent that sends `tick' (or \Ltick), and right is the agent
that replies `tock' (or \Rtick).

Boldface vectors $\bf H$ denote intent-acknowledgment 3-vector
message headers.
Message headers promise to signal intent of a transmission in one of three cases $(\emptyset,+,-)$.
In addition, they encode the intended recipient of the message
(one bit for left or right $(L,R)$), because a message may conceivably be misread
or reflected back by the physical environment to the initiator.
So the header data need
\beq
\text{direction} \otimes \text{intent} \rightarrow (L,R) \otimes (I,S,R)
\eeq
These six cases can be represented in three bits, or in the components of a 3-vector: 
\beq
\bf H &=& \left(
\begin{array}{c}
\text{Intent}\\
\text{LR-Ack}
\end{array}
\right)\\
~\nonumber\\
~\nonumber\\
&\stackrel{\text{promise}}{=}& \left(
\begin{array}{c}
\vec{(\emptyset,+,-)}\\
\text{(L,R)}
\end{array}
\right)\\
~\nonumber\\
~\nonumber\\
&\stackrel{\text{bit rep.}}{=}& 
\left\{
\begin{array}{c}
\left(\begin{array}{c} 0/1\\0/1\end{array}\right),
\left(\begin{array}{c} 1\\0\end{array}\right),
\left(\begin{array}{c} 0\\1\end{array}\right)\\
\{0,1\}
\end{array}
\right\}\nonumber\\
\eeq

Note that the encoding of idling could be symmetrical with either 0s or 1s.
The final choice takes one of each, as is explained below.
A link transmission header, in either left or right direction may take one of the following
forms:

\beq
\text{\sc Label} &\leftrightarrow& \text{\sc 2-intent} \oplus \text{\sc 1-direction}\nonumber\\
\hline~\nonumber\\
\Ltick &\leftrightarrow&  \emptyset  \oplus L2R \; \leftrightarrow \left(\begin{array}{c} 0\\0\\0\end{array}\right)\label{sup1}\\
\Rtick &\leftrightarrow&  \overline \emptyset\oplus R2L \; \leftrightarrow \left(\begin{array}{c} 1\\1\\1\end{array}\right)\label{tock}\\
\Lteck &\leftrightarrow&  +\oplus L2R \;  \leftrightarrow \left(\begin{array}{c} 1\\0\\0\end{array}\right)\\
\Rteck &\leftrightarrow&  + \oplus R2L \;  \leftrightarrow \left(\begin{array}{c} 1\\0\\1\end{array}\right)\\
\Ltack &\leftrightarrow&  -\oplus L2R \; \leftrightarrow \left(\begin{array}{c} 0\\1\\0\end{array}\right)\\
\Rtack &\leftrightarrow&  -\oplus R2L \; \leftrightarrow \left(\begin{array}{c} 0\\1\\1\end{array}\right)\label{sup2}
\eeq

When no messages are being sent, and the boundary conditions on the link
from the parent cells are only symmetrical over whole numbers of cycles,
but we can't see this at the link layer, which remains a ticking clock
of modulo 2, skipping or treading water.
We thus write the components
as vectors that encode the {\em orientation} of the message relative
to the endpoints.  Although the promises reach a promise equilibrium,
the dynamical realization has a handedness in virtue of the broken
symmetry. This originates essentially from the order of preconditions
imposed by the cell agents, in the form of boundary constraints on the
dynamics of the link\footnote{Note that the ability to distinguish left
  from right depends on the timescale of measurement, i.e. on whether
  one measures time intervals preferentially in even ($2n$) or odd
  ($2n+1$) numbers of ticks. Since taking a large number $n
  \rightarrow \infty$ of ticks makes this distinction irrelevant, one
  can say that behaviour that is, on average, the same in both
  directions is undirected on average.  This limit is what is
  important to the long term behaviours of a link, while the short
  term fluctuations may be directional. This means we can send short
  term packets comprising messages, intentionally, in either direction
  without sacrificing long term coherence.}.

\subsection{Packet header 3-vectors:  (in/out) complement}

As data circulate between sender
(intender) and receiver (acknowledger) roles, two parts of information
remain anti-correlated as complement reflections of one another.
The purpose of entanglement is that it pushes intent to the far
extrema of the link, i.e. to the edges where boundary conditions are
determined, and away from the places where noise can enter en route.
This has the effect of making noise less of an issue.
As long as we maintain entanglement, at both link and transmission
layers, we effectively know the state of the other agent, because the
states of the sender and receiver are mirror images within each channel of intent.

In practice, the intention to send (without the payload) is merely a
direction vector L2R or R2L indicating who is sender and who is
recipient: a simple auto-addressing scheme for a point-to-point link.
Thus it always points in the direction of travel, while the
acknowledgment component continues to tick-tock. When the tick or
tock matches the symmetry breaking of left and right, the receiver
register knows that it should accept or ignore the content on the
line\footnote{On short cables with only a single duplex wire,
  transmission is basically instantaneous to both ends of the cable,
  so it may not be obvious, depending on the implementation.}.

Let's define the bar operation to be the simple one's complement of each
component. Then we see that simple reflection implies\footnote{Note, that the encoding of \tick and \tock don't need to
satisfy this property, because the natural choice from a purely information perspective
might be $\tock = R2L \oplus I$ (rather than $\overline I$), however
by assuring the complement explicitly in (\ref{tock}) we can maintain validation of the entanglement explicitly on
every reflection.}:
\beq
\overline \tick &=& \tock\\
\overline \tock &=& \tick\\
\overline {\Lteck} &=& \Rtack\\
\overline {\Rteck} &=& \Ltack.
\eeq
Note, to fully expose the L,R symmetry, we could 
further regularize the notation and define:
\beq
 \tick &=& \Ltick\\
 \tock &=& \Rtick\\
\overline \Ltick &=& \Rtick\\
\overline \Rtick &=& \Ltick,
\eeq
making the idling phase and the transient message
phase formally similar. Indeed, with this notation, there is a simple
promise correspondence:
\beq
\tick &\lrarrow& \emptyset\\
\teck & \lrarrow& +\\
\tack &\lrarrow& -
\eeq
This complementarity property is entangled with the entanglement!
In other word, it holds as long as the mutual promises are kept. 
If it fails to be true for whatever reason, then 
entanglement has been lost and an error has occurred.
The link must then be re-established for continuation of a message.

In other words, `what I know about me' (local state)
(intent) and `what I know about you' are compared and validated by
this header property alone, as long as both agents keep the same calibrated
promises.

\subsection{Steady state interaction}\label{simple}

\begin{figure}[ht]
\begin{center}
\includegraphics[width=7.5cm]{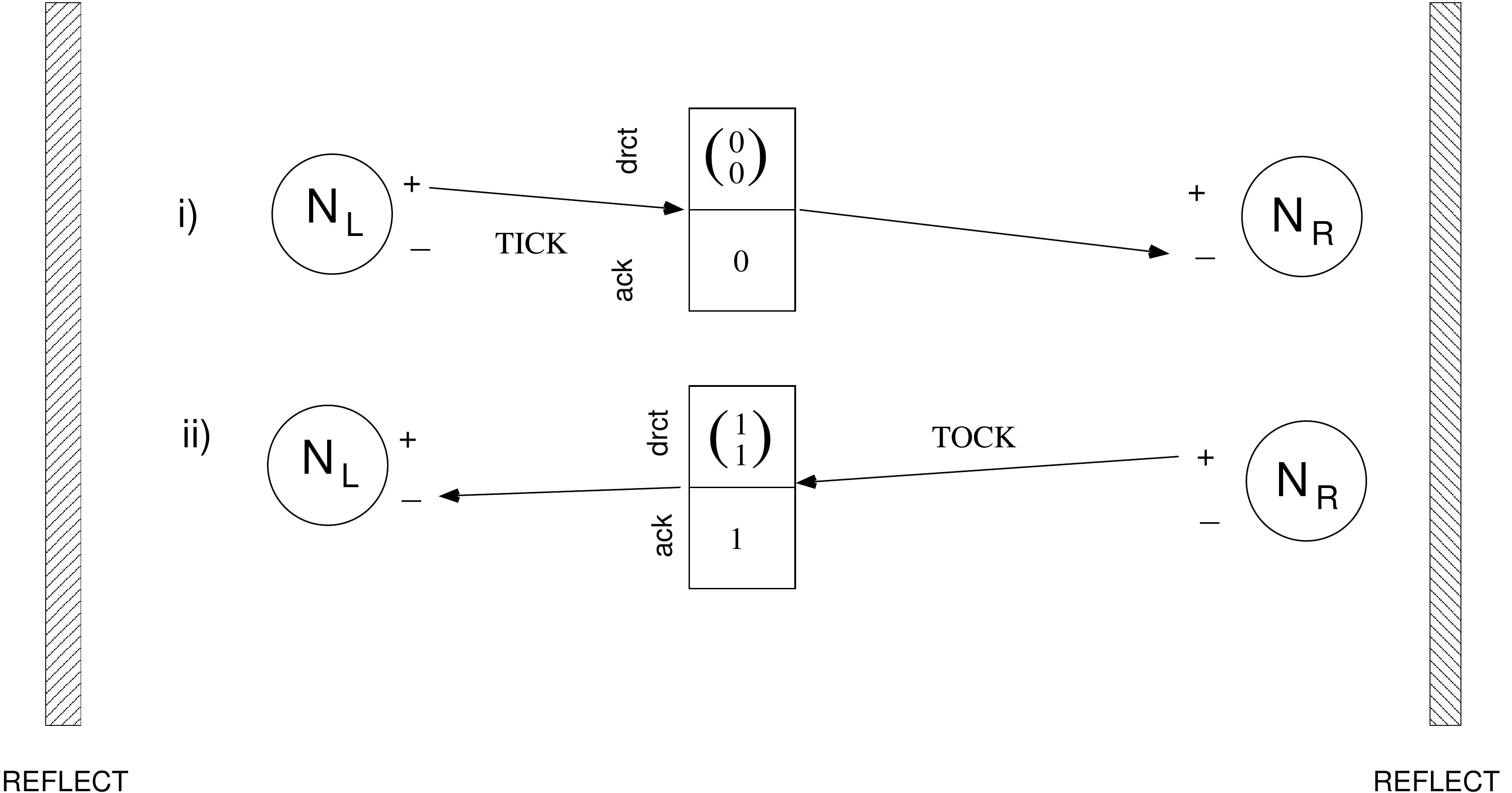}
\caption{\small The simplest kind of exchange is the reflection of a `hot potato' token
between the end points of the link, like a photon reflecting back and forth, `treading water'
and going nowhere. At each end, there is a parity reversal represented by the complement operation.\label{ticktock}}
\end{center}
\end{figure}

\begin{figure}[ht]
\begin{center}
\includegraphics[width=7.5cm]{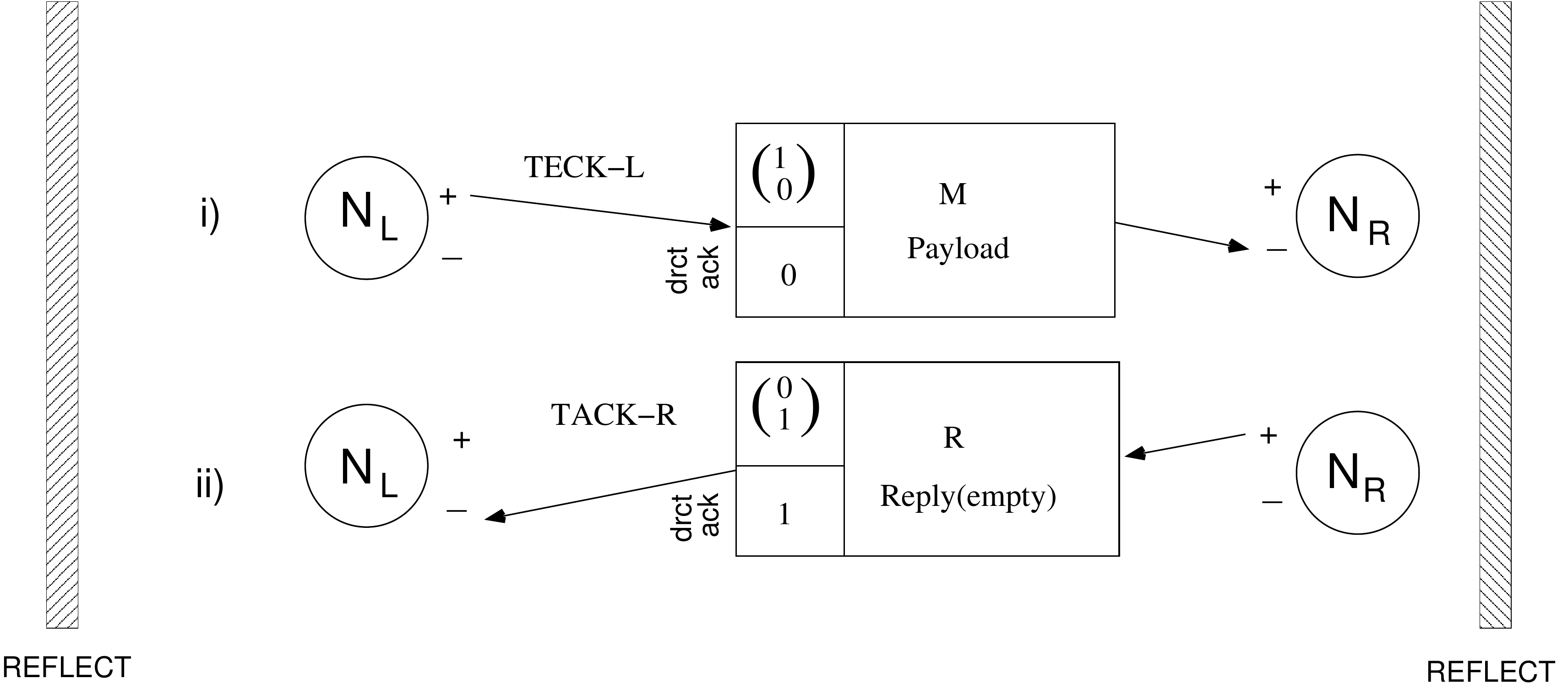}
\caption{\small {\bf Left to right transmission}: In order to send a message the `treading water' phase needs to be broken
by an intentional boundary condition imposed at one of the ends of the link.
The exchange of a message, which may be compared to figure \ref{ENTT} 
from left to right breaks thus from the directionless tick-tock phase.
\label{L2R}}
\end{center}
\end{figure}

\begin{figure}[ht]
\begin{center}
\includegraphics[width=7.5cm]{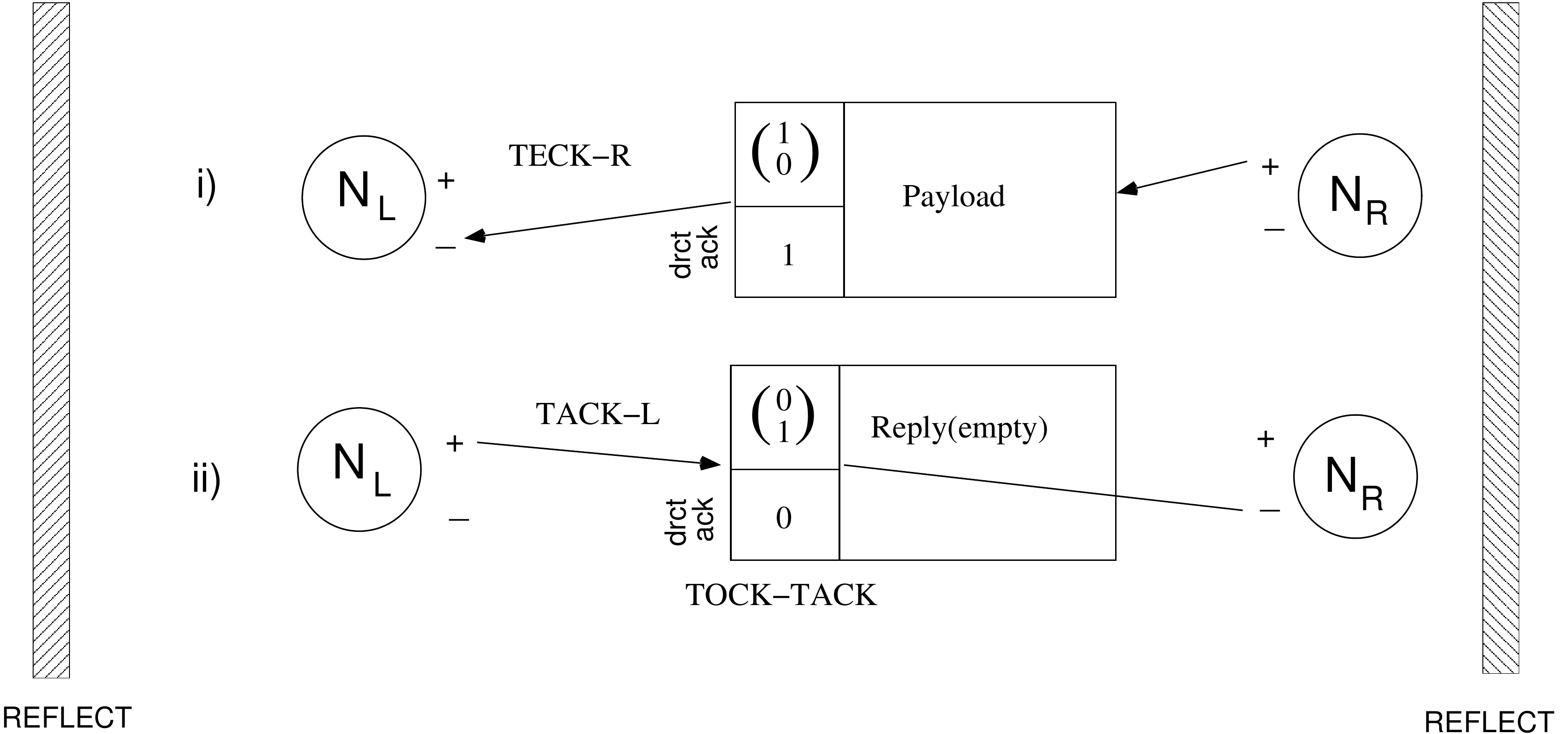}
\caption{\small {\bf Right to left transmission}: In order to send a message the `treading water' phase needs to be broken
by an intentional boundary condition imposed at one of the ends of the link.
The exchange of a message, which may be compared to figure \ref{ENTTR} 
from left to right breaks thus from the directionless tick-tock phase.\label{R2L}}
\end{center}
\end{figure}

\begin{figure}[ht]
\begin{center}
\includegraphics[width=7.5cm]{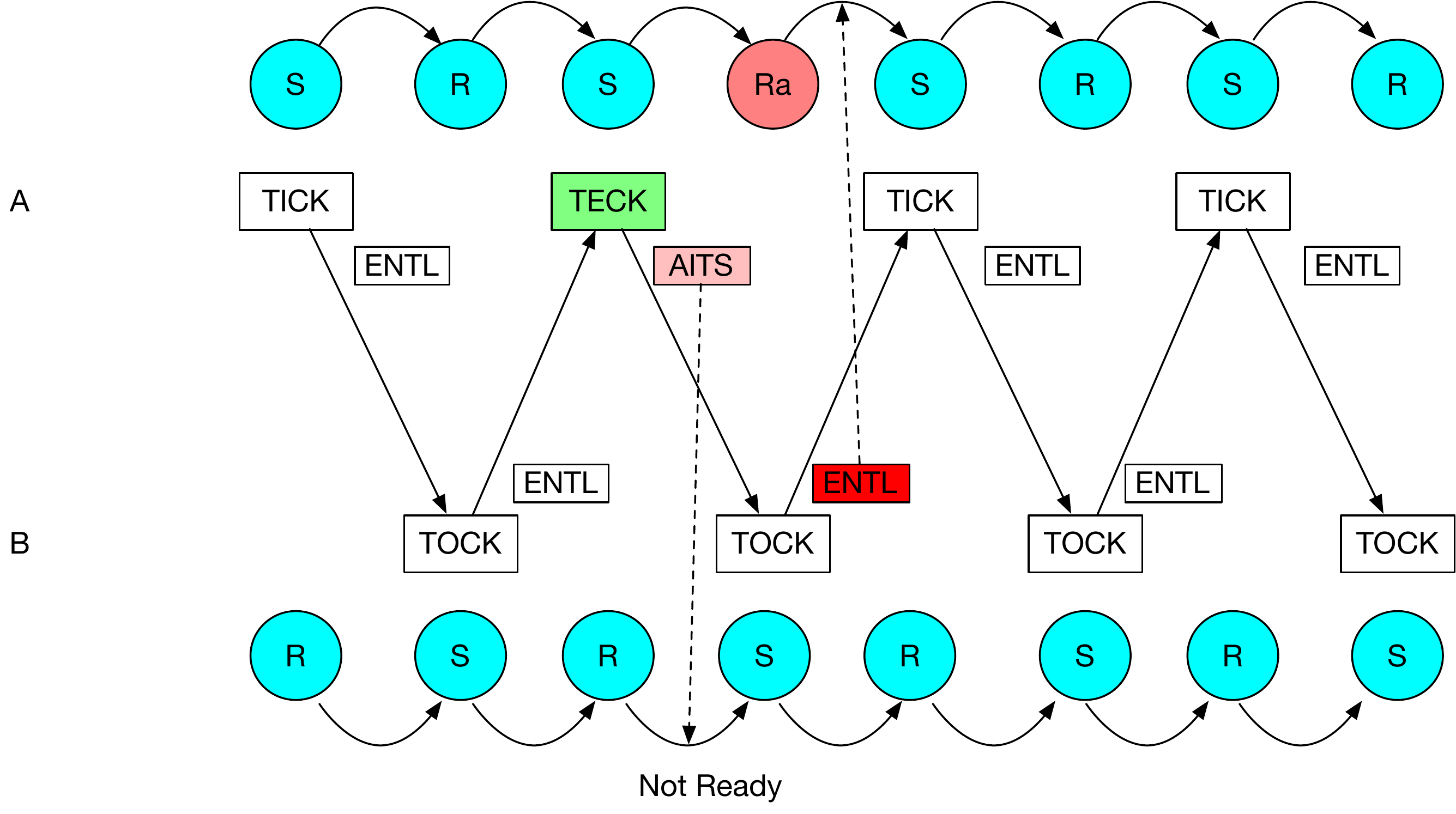}
\caption{\small Schematic L2R or A2B transfer, with interior states of endpoint agents indicated.\label{canvas6}}
\end{center}
\end{figure}

The simplest version of the interaction has a single alphabet, and
symmetrical idling (figure \ref{ticktock}) and asymmetrical message
(figures \ref{L2R} and \ref{canvas6} for left to right, and figure
\ref{R2L} for right to left) modes of transfer.


The promises we intend to keep here:
\begin{itemize}
\item To maintain entanglement of the link, ensuring a sense of shared
  time on the interior of the link. This can be used to enable ordered
  delivery on a single serial connection (prove this) within the scope of the
entanglement.
\item To utilize a consistent tick-tock encoding and exchange
  throughout all message passing (a single `carrier wave', in contrast to the two
mode carrier encoding in the ENTT, ENTL version).
\item No promises can be made about ordering over parallel channels.
\item The promises made by the payload are undefined.
\end{itemize}

Figure \ref{L2R} shows a packet sent from left to right by jumping
onto a clear \tick package. Figure \ref{R2L} shows a packet sent from
right to left by jumping onto a clear \tock package. In both cases,
the link only needs to detect whether the registers signal a
directionless state (with equal components) for both $I$ and $A$
fields to know that any previous transmission has ended, and that the
network interface is clear to send. If the $I$ field contains a
direction vector (with unequal components) then it is busy.

Since the send $N_+$ and receive $N_-$ registers of a network interface
are connected only by their being part of the network interface  itself,
the interface needs to promise the circumstances or preconditions under which
the link would be held up to wait for acceptance of a delivered packet, or whether
some complicating buffering would be added that admits only partial delivery.
The network interface needs to promise these details as part of its 
self-specification.

\begin{figure}[ht]
\begin{center}
\includegraphics[width=7.5cm]{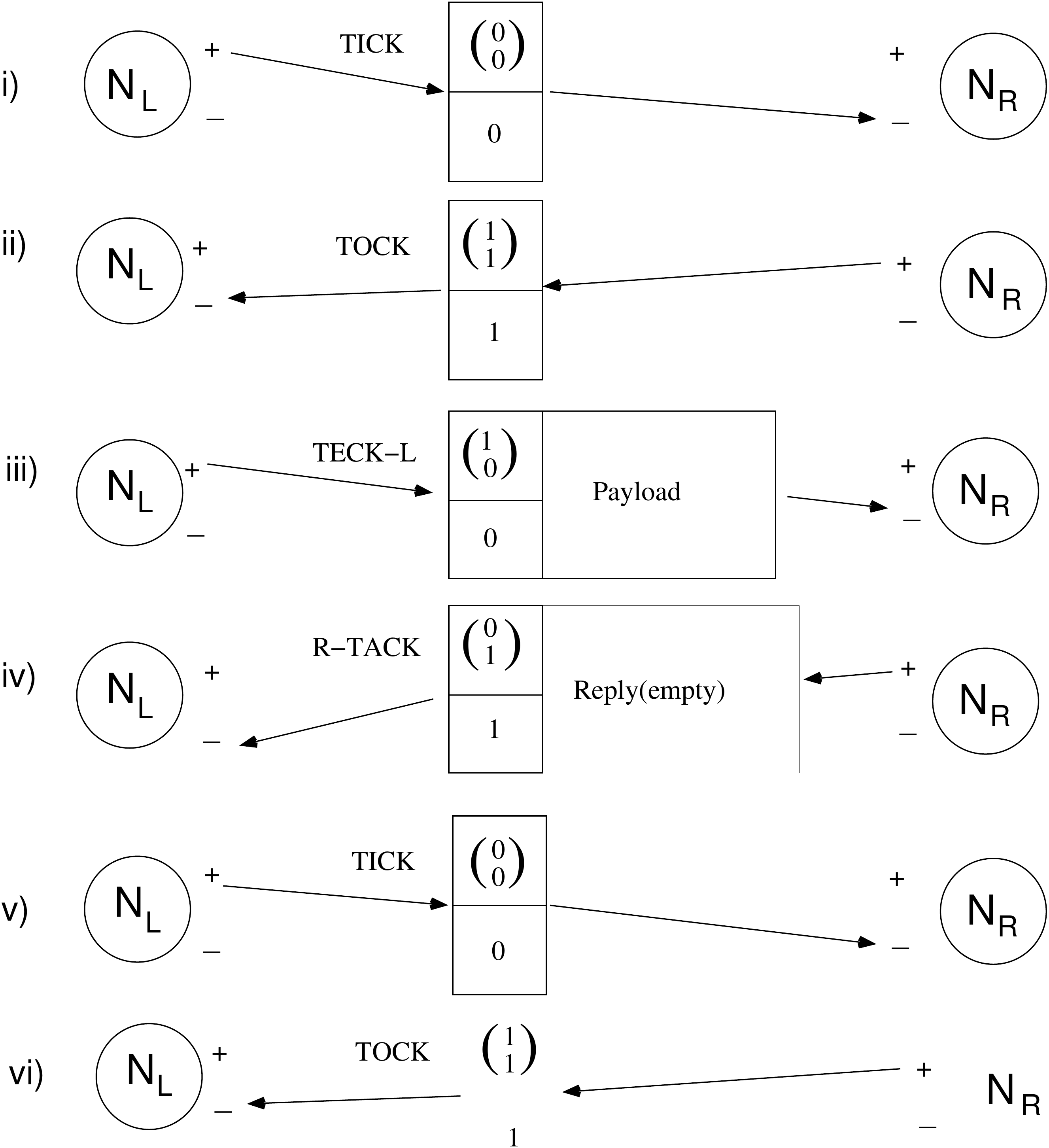}
\caption{\small A left to right transfer, with injection of a packet and acknowledgment, between pendulum phases \cite{classent}.\label{ENTT}}
\end{center}
\end{figure}

\begin{figure}[ht]
\begin{center}
\includegraphics[width=7.5cm]{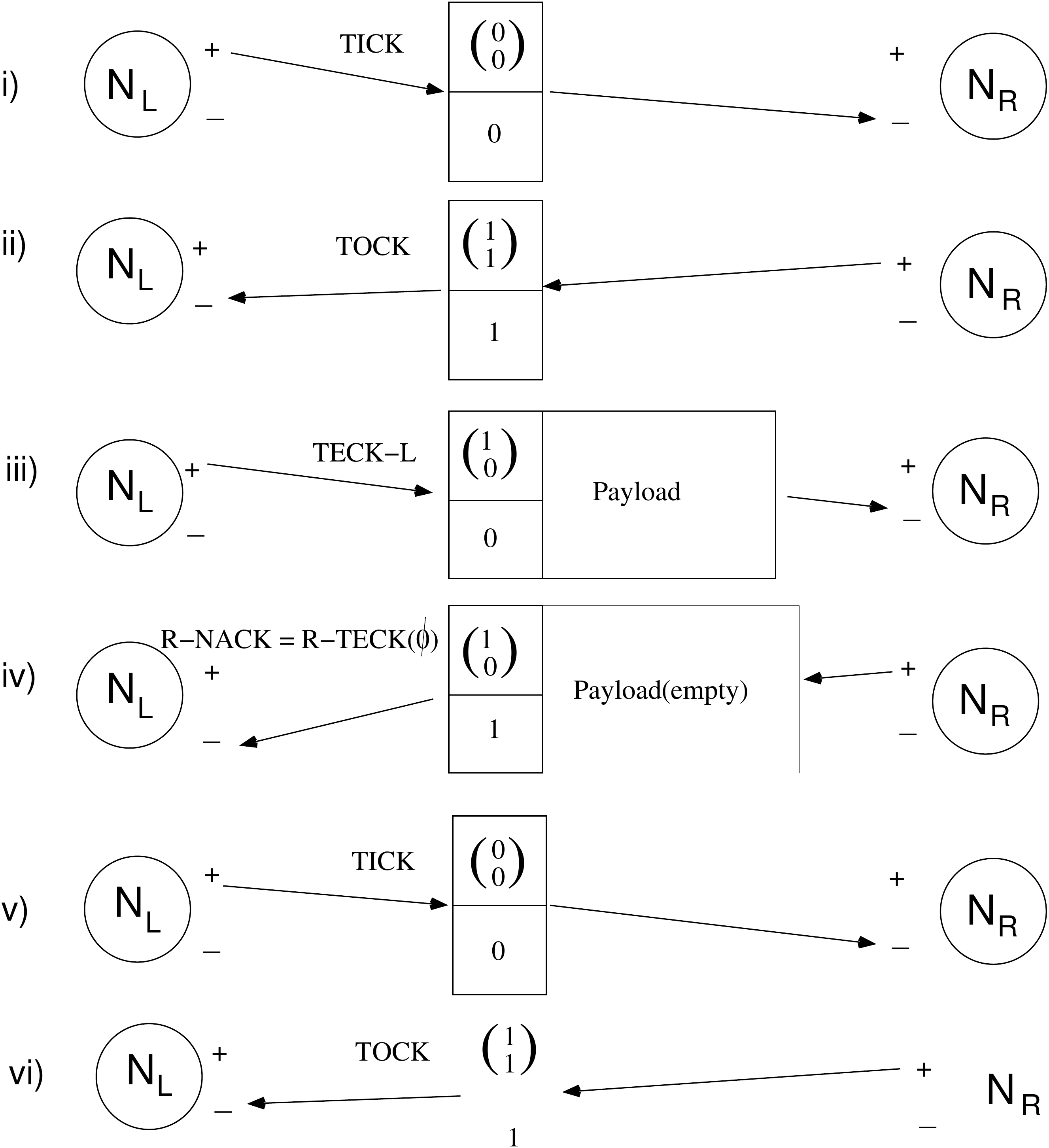}
\caption{\small A left to right transfer, with non-accepted payload, returning \nack instead of \tack..\label{reject}}
\end{center}
\end{figure}

\begin{figure}[ht]
\begin{center}
\includegraphics[width=7.5cm]{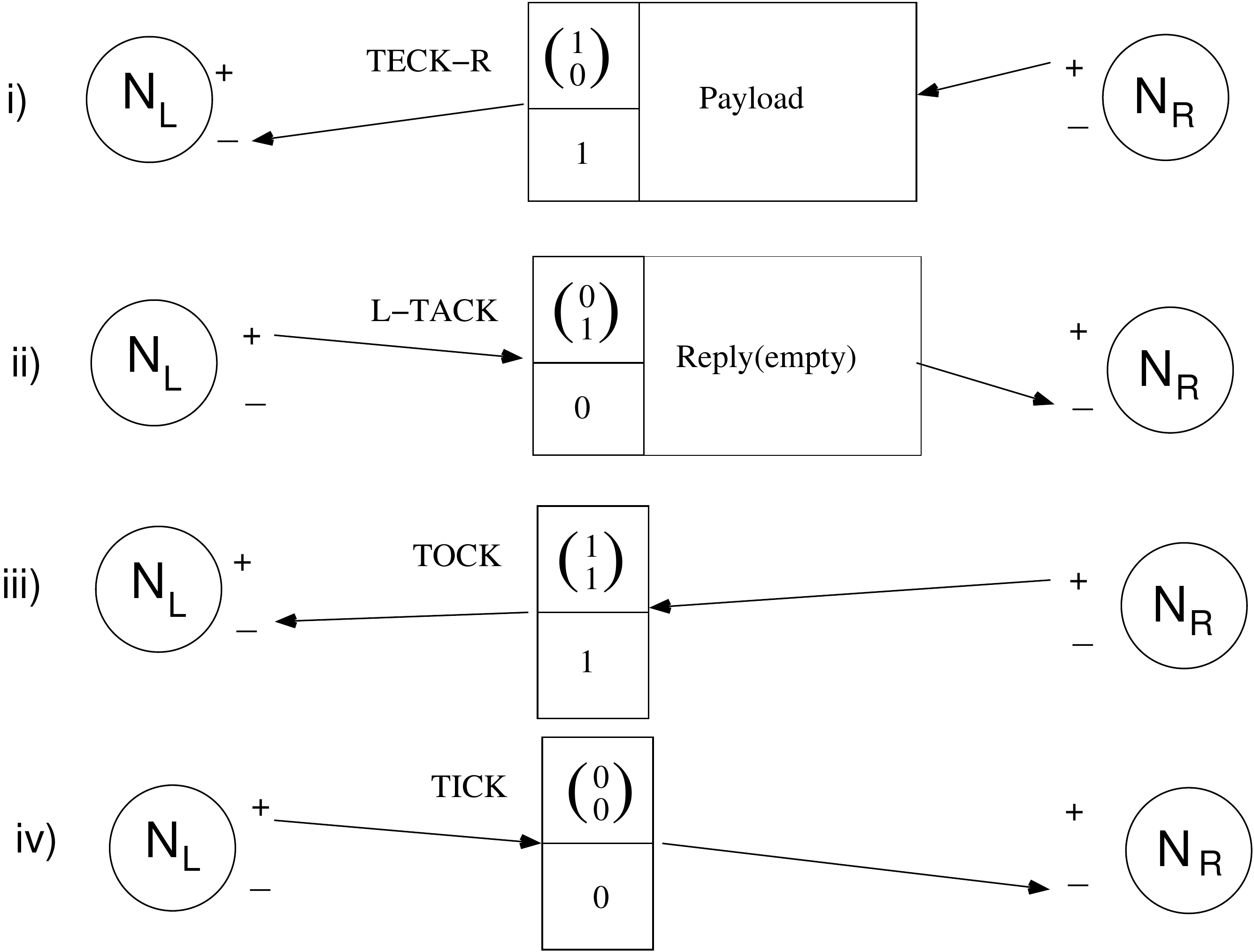}
\caption{\small A right to left transfer, with switch from pendulum to transfer phase.\label{ENTTR}}
\end{center}
\end{figure}

\end{document}